\documentclass[aps,prb,amsmath,amssymb,nobibnotes,longbibliography,superscriptaddress,twocolumn]{revtex4-2}

\usepackage{t1enc}
\usepackage[utf8]{inputenc}
\usepackage{amsmath}

\usepackage{graphicx,bm,amsmath,amssymb,natbib,color}
\usepackage[unicode=true, colorlinks=true, citecolor={blue!80!black}, urlcolor={blue!50!black}, linkcolor = {blue!80!black}]{hyperref}
\usepackage{physics}
\usepackage{bm}
\usepackage{bbm}
\usepackage[dvipsnames]{xcolor}
\usepackage{soul}

\usepackage[english]{babel}
\begin{document}
\title{Quantum circuits with multiterminal Josephson-Andreev junctions}

\newcommand{\affA}{\affiliation{Departamento de F\'{\i}sica Te\'orica de la Materia Condensada, \mbox{Condensed Matter Physics Center (IFIMAC)} and Instituto Nicol\'as Cabrera, Universidad Aut\'onoma de Madrid, 28049 Madrid, Spain}}
\newcommand{\affC}{\affiliation{Centro At\'omico Bariloche and Instituto Balseiro, CNEA, CONICET, 8400 San Carlos de Bariloche, R\'io Negro, Argentina}}

\author{F. J. \surname{Matute-Ca\~nadas}}
\affA
\author{L. Tosi}
\affC
\author{A. \surname{Levy Yeyati}}
\email[Corresponding author: ]{a.l.yeyati@uam.es}
\affA

\date{\today}

\begin{abstract}
We explore superconducting quantum circuits where several leads are simultaneously connected beyond the tunneling regime, such that the fermionic structure of Andreev bound states in the resulting multiterminal Josephson junction influences the states of the full circuit. 
Using a simple model of single channel contacts and a single level in the middle region, we discuss different circuit configurations where the leads are islands with finite capacitance and/or form loops with finite inductance. We find situations of practical interest where the circuits can be used to define noise protected qubits, which map to the bifluxon and $0{-}\pi$ qubits in the tunneling regime. We also point out the subtleties of the gauge choice for a proper description of these quantum circuits dynamics.
\end{abstract}
\maketitle

\section{Introduction}
Josephson junctions are essential ingredients for turning quantum circuits into artificial atoms \cite{Clarke1988,Blais2021}. When the degrees of freedom describing the circuit are quantized, the non-linear non-dissipative Josephson element introduces the anharmonicity in the potential energy that is required to isolate a set of levels for their use as a computational basis. A conventional tunnel junction is characterized by its charging and Josephson energies, determined, respectively by its capacitance and its critical current, which depend on the tunnel barrier and the junction geometry. This description of the junction as a non-linear inductive lumped element is justified when the energy corresponding to the fermionic excitations of the junction itself is large compared to the lowest circuit levels and, thus, only bosonic excitations are considered.

\begin{figure*}[t]
    \centering
\includegraphics[width=1.97\columnwidth]{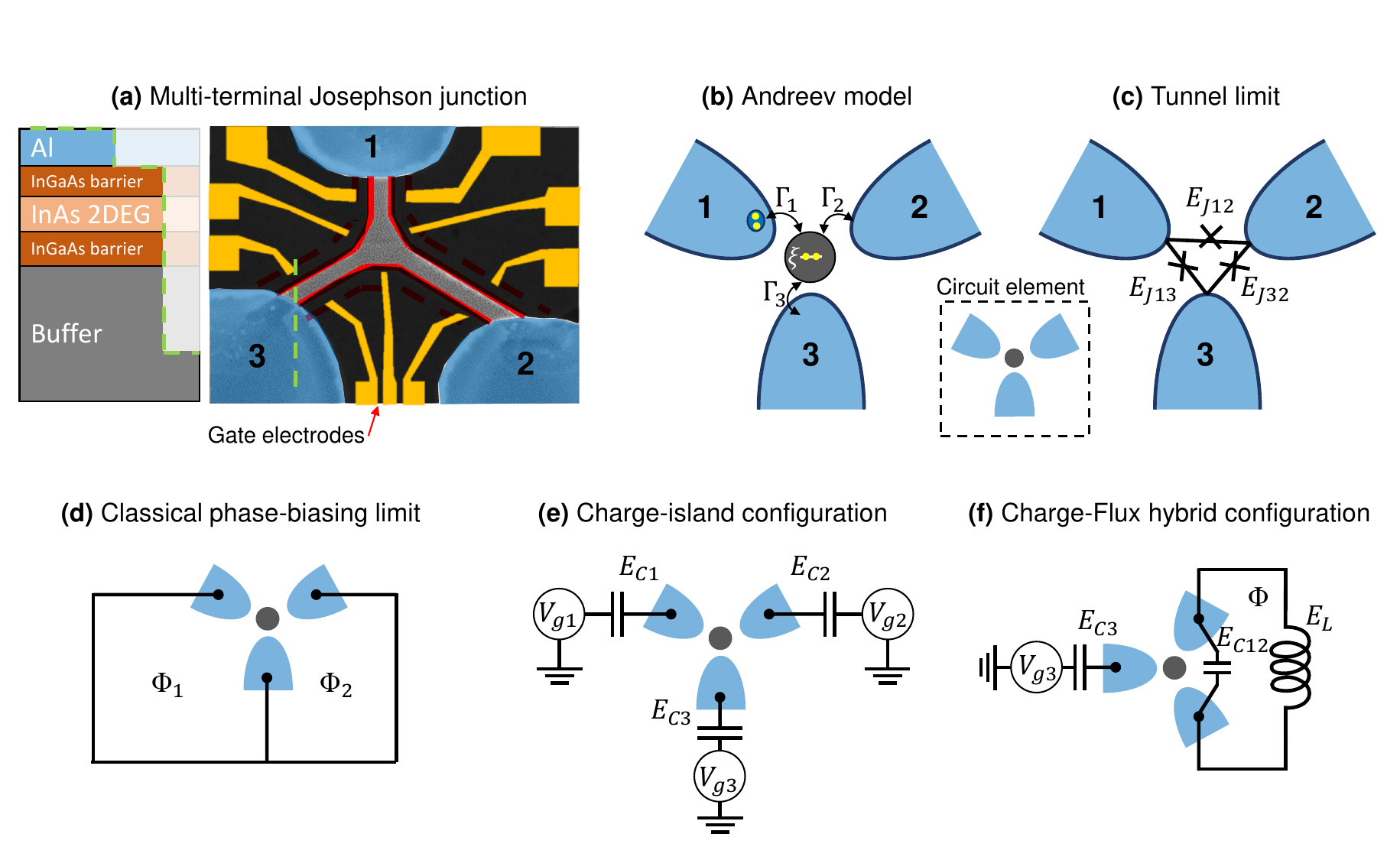}
    \caption{(a) Possible implementation of the trijunction in a hybrid superconductor-semiconductor heterostructure. An aluminium film epitaxially grown on top of a 2DEG (pale orange) induce superconductivity by proximity. The 2DEG can be patterned by a defining a mesa and the junctions are created by etching the Al in some regions (indicated with red lines). The gate electrodes confine the electrons of the 2DEG in the center and control the coupling with each lead. (b,c) Models describing the trijunction. The circuit element indicated in a dashed box represents both limits in (d-f). In (b), the middle region is relevant to describe how electrons move between leads. In this work, we use a single level quantum dot in the infinite gap regime at even parity, such that the relevant processes involve the exchange of Cooper pairs between the dot and the leads. In (c), the system is described by effective tunneling processes of single Cooper pairs between the leads -equivalent to 3 tunnel Josephson junctions forming a triangle-, and may reproduce the model in (b) under certain limits. (d-f) Three circuits we consider with the trijunction element. In (d), the phase is externally fixed by the external fluxes $\Phi_{1,2}$. In (e), the trijunction connects three superconducting islands with charging energies $E_{C\nu}$. In (f), two leads form a loop with inductive energy $E_L$, threaded by flux $\Phi$, and the third lead is a superconducting island. \label{fig:sketch}}
\end{figure*}

More generally, from a mesoscopic perspective, a tunnel junction is just a particular case of a weak link between two superconductors \cite{nazarov_blanter_2009}. The coupling being weak, bound states with phase-dependent energies form to accommodate the phase difference between both terminals \cite{Kulik1969,Furusaki1991,Beenakker1991}. Solving the energy spectrum of the so-called Andreev states of this \textit{SXS} device means the diagonalization of the Bogoliubov-de Gennes (BdG) Hamiltonian, i.e. a fermionic problem that requires a microscopic model for the \textit{X} region.
In this BdG description, the phase is a parameter determined by the circuit where the weak link is embedded, an approach justified when the phase fluctuations are small and the energy corresponding to the excited circuit levels is large compared to the Andreev states.

The hybrid situation, involving a mesoscopic junction where the phase has to be considered as an operator associated with the transfer of charge has acquired interest with the recent realizations of quantum circuits containing Josephson junctions made out of InAs-nanowires and two-dimensional electron gases (2DEGs) \cite{Bargerbos2020,PitaVidal2022,Kringhoj2018,Kringhoj2020, Coraiola2023_2phases}. These junctions may operate away from the tunneling regime and are better described in terms of high transmission conduction channels or with a quantum dot model. The number of channels, the barriers and chemical potentials can be tuned using gates, thus enabling the realization of devices where the control of the junction parameters is critical.
Previous works have already considered this \textit{Josephson-Andreev} regime at different degrees of approximation, both in situations with topological leads, where the fermionic levels correspond to Majorana states  \cite{Fu2010,vanHeck2011,Karki2023,Keselman2019,Avila2020,Avila2020prr,Pino2023}, and with trivial Andreev states \cite{Bretheau2014,Keselman2019,Vakhtel2023,Caceres2022,Vakhtel2023flux}.

These devices open up the possibility to design hybrid modes combining the fermionic structure of the weak link with the bosonic excitations associated to the electromagnetic collective modes. This is in line with nowadays efforts towards the physical realization of {\it protected} qubits, based on the concept of a disjoint support of the qubit wavefunctions, which enjoy protection against decoherence from one or more noise sources \cite{gyenis2021, Calzona2022,danon2021,Doucot2012}. Such a strategy has been pursued using multimode circuits with conventional junctions \cite{Brooks2013, Dempster2014, Groszkowski2018, Paolo2019, Gyenis2021_exp0pi, Kalashnikov2020, Bell2016, Doucot2002, Smith2020, Protopopov2004, Protopopov2006}, mesoscopic junctions where the effect of the Andreev structure is approximated with a modified Josephson potential \cite{Larsen2020, Aguado2020, Casparis2018,Chirolli2021,Maiani2022,Schrade2022,Guo2023,Patel2023}, and mesoscopic junctions where the treatment of the full Andreev structure is essential, as is the case of a quantum-dot with a level in resonance with the Fermi level of the superconducting leads \cite{Bargerbos2020,Kringhoj2020, Vakhtel2023,Caceres2022,Vakhtel2023flux}.

In this work we explore the properties of multiterminal Josephson-Andreev junctions when immersed in different kinds of quantum circuits (see Fig. \ref{fig:sketch}). These are unique mesoscopic components where the weak-link connects several superconducting leads that have become experimentally accessible in proximitized 2D structures \cite{Pankratova2020,Graziano2020}. Since there are several phase difference variables which mimic the $k$ space in a solid, there has been considerable interest in multiterminal Josephson devices driven by the possibility of engineering topologically protected singularities in their spectrum \cite{Riwar2016,Fatemi2021,PeraltaGavensky_2023, PeraltaGavensky2018, Teshler2023,Klees2020,Klees2021,Yokoyama2015,vanHeck2014,Meyer2017,Xie2017,Xie2018,Peyruchat2021,Mukhopadhyay2023, Zalom2023} and, broadly, other applications that arise from the introduction of additional leads \cite{Melin2017,Deb2018,Melin2021,Melo2022,Melin2023,Cayao2023,Pankratova2020,Padurariu2015,Zazunov2017,Amin2001,Amin2002,Graziano2022,Gupta2023,Melin2023nov,Prosko2023,ArayaDay2023,Coraiola2023,Ohnmacht2023,Coraiola2023_4diode}. We find that, just at the level of a triterminal junction [Fig. \ref{fig:sketch}a], the presence of a third lead introduces an additional degree of freedom that allows to define protected qubits.

At a more fundamental level, quantum circuits with Josephson-Andreev junctions pose basic questions on the proper quantization rules that determine the system Hamiltonian in a general, time dependent situation. For instance, if a time dependent external flux constrains the phases of some circuit elements, the phase drop distribution between and within these elements is not a gauge freedom as in the static situation. The actual drop has to be determined by the electromagnetic field established in the circuit, which depends on its structure and geometry \cite{You-Sauls-Koch2019,Riwar2022,Bryon2023,Kenawy2022,Kenawy2023} and on the currents spatial distribution \cite{Zazunov2005,Ivanov1999,Ambegaokar1982,Eckern1984,Pham2023,Grankin2023,Kurilovich2021}. In addition, the circuit variables bear a different nature depending on the configuration of the circuit, leading to continuous or discrete charge variables \cite{Devoret2021,Sonin2022,Osborne2023,Crescini2023,Murani2020,Schon1988,Apenko1989,Schon1990,Riwar2021,Koliofoti2022,Herrig2023}. In the present work we analyze this problem for different circuit configurations and show its impact on the relaxation rates.

The work is organized as follows: in Section \ref{sec:phase_bias}, we introduce possible experimental implementations and discuss the modeling of the junction with a single level between the leads [Fig. \ref{fig:sketch}b], its embedding in a phase biased configuration [Fig. \ref{fig:sketch}d] and the effect of the phase drop distribution on the relaxation rates.
In Section \ref{sec:island}, we consider a circuit where the leads are superconducting islands [Fig. \ref{fig:sketch}e] and discuss a transmon-like regime that hosts two almost degenerate states. In section \ref{sec:loop}, we connect two leads in a loop with finite inductance with a third one, which is an island [Fig. \ref{fig:sketch}f], and discuss a regime simultaneously protected against charge and flux noise. In section \ref{sec:cotunneling}, we describe the tunnel limit of a multiterminal junction [Fig. \ref{fig:sketch}c] and its connection to protected multimode circuits designed with several conventional elements. In section \ref{sec:overview}, we summarize and compare the main features of the analyzed circuits, discuss the effect of quasiparticle poisoning and elaborate on the analysis of decoherence for feasible implementations. We finish with some conclusions in section \ref{sec:conclusions}. Additional details can be found in the Appendixes. 

\section{Modeling, phase biased configuration and gauge choice}
\label{sec:phase_bias}

The basic element that we consider as a building block for quantum circuits is the Josephson-Andreev trijunction. As illustrated in Fig. \ref{fig:sketch}(a), it consists of a normal central region coupled to three superconducting leads. Such a device can be implemented on a proximitized 2DEG defined, for instance, on a Al/InAs heterostructure \cite{Shabani2016,Coraiola2023_2phases,Baumgartner2021,Baumgartner2021nano,Costa2023}. We assume that the charge density in the normal region can be controlled by gates and that a situation where a few or just one conduction channel on each terminal can be reached. In such situation the Josephson coupling between the leads is mediated by a few Andreev bound states, sensitive to the phase on each superconducting contact. Notice that in the configuration of Fig. \ref{fig:sketch}(d) two well defined phase differences between the leads can be controlled by external fluxes $\Phi_{1,2}$ through the corresponding loops, as indicated in the figure. Within this section, we assume ideal phase biasing, i.e. loops without inductance and negligible charging energy.

As commented in the introduction, determining the Andreev spectrum for such geometry requires solving the BdG equations with the corresponding boundary conditions for the superconducting phases. This is, in general, a formidable task that can be undertaken using different models and numerical techniques. For the purpose of the present work, however, we adopt several simplifying assumptions that enable a tractable but still realistic description of the device.

Assuming a single channel per lead and that the dimensions of the normal region are short compared to the superconducting coherence length, we describe it in terms of a single level model. We denote by $d^{\dagger}_{\sigma}$ the operator creating an electron with spin $\sigma$ in that level. In addition, we describe the leads in the so-called infinite-gap limit, which allows us to include their effect in the normal region as a frequency independent pairing self-energy \cite{Affleck2000,Meng2009,MartinRodero2011}. The effective Hamiltonian for the trijunction is given by (see Appendix \ref{ap:gauge})
\begin{equation}
    H^{\rm eff} = \sum_{\sigma} \epsilon n_{\sigma} + U n_{\uparrow}n_{\downarrow} + \sum_{\nu=1}^3 \left[\Gamma_{\nu} e^{-i\phi_{\nu}} d^{\dagger}_{\uparrow}d^{\dagger}_{\downarrow}+
    \mbox{h.c.} \right]\;,
    \label{Heff}
\end{equation}
where $\epsilon$ is the position of the central level referred to the leads chemical potential, $U$ is its charging energy, and $\Gamma_{\nu}$ is the tunneling rate to the lead $\nu$, which has a phase $\phi_{\nu}$. 

\begin{figure}[t]
    \centering
    \includegraphics[width=1\linewidth]{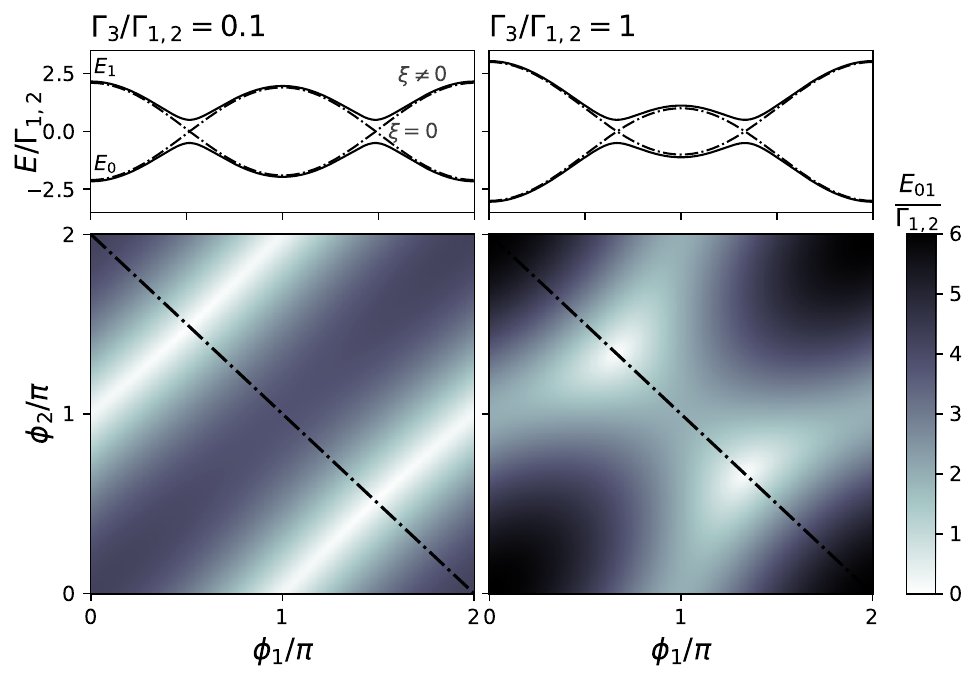}
    \caption{Bottom: transition energy $E_{01}$ versus the phase differences $\phi_{1,2}$ for $\Gamma_1{=}\Gamma_2$ and different $\Gamma_3$. At low coupling $\Gamma_3$ (left), the system roughly behaves as a junction between $1$ and $2$, and depends on the phase difference $\phi_1-\phi_2$ between them. At comparable couplings $\Gamma_{1,2,3}$ (right), the transition landscape has a set of local minima. The level position is placed at the symmetry point $\epsilon{=}-U/2$, meaning $\xi{=}0$. Top: energies $E_{0,1}$ (up to a shift) of the Andreev states along the trajectories highlighted in the maps, with solid (dash-dotted) lines for $\xi{=}0.5\Gamma_{1,2}$ ($\xi{=}0$). \label{fig:PhaseBiasedTrijunction}}
\end{figure}

Such a model is characterized by a phase diagram with alternating even and odd parity ground states as a function of the model parameters \cite{MartinRodero2011,Fatemi2022,Bargerbos2022,Teshler2023,Baran2023,Zalom2023}. However, as in the present work we focus on the case of widely open channels with significant Josephson coupling, we assume that the trijunction remains always within the even sector, where $H^{\textrm{eff}}$ is projected as
\begin{equation}
H^{\textrm{eff}}_{\textrm{even}} = -\xi\tau_z + \xi + \sum_\nu \Gamma_\nu (\cos{\phi_\nu}\tau_x-\sin{\phi_\nu}\tau_y),
\label{Heffeven}
\end{equation}
with $\xi=\epsilon+U/2$ and $\tau_i$ the Pauli matrices acting on the space $\{\ket{0}_d,\ket{\uparrow\downarrow}_d\}$. The ground and excited states have energies 
\begin{equation}
E_{0,1}(\{\phi_{\nu}\}) = -\xi \mp \sqrt{ \xi^2 + \bigg| \sum_{\nu} \Gamma_{\nu} e^{-i\phi_{\nu}}\bigg|^2 } \;.
\label{spectrum}
\end{equation}

Let us note that, while $H^{\rm eff}$ in Eq. (\ref{Heff}) does depend on the phase gauge choice, its spectrum (\ref{spectrum}) is only sensitive to the phase differences so for simplicity we take $\phi_3=0$. The transition energy $E_{01}=E_1{-}E_0$ is shown in Fig. \ref{fig:PhaseBiasedTrijunction} (bottom) for $\Gamma_1=\Gamma_2$ and different ratios $\Gamma_3/\Gamma_{1,2}$. When lead $3$ is weakly coupled (left panel), the energy depends mainly on the phase difference between the two strongly coupled leads, $\phi_{12}{=}\phi_1{-}\phi_2$, which at $\pi$ corresponds to a minimum in $E_{01}$. Indeed, in the limit $\Gamma_3{=}0$ we recover the familiar expression $E_{0,1}(\phi_{12}) + \xi = \mp \sqrt{\xi^2+\Gamma^2}\sqrt{1-T\sin^2 (\phi_{12}/2)}$, where $T=(\Gamma^2-\delta\Gamma^2)/(\xi^2+\Gamma^2)$ is the transparency of the junction, with $\Gamma, \delta\Gamma{=}\Gamma_1{\pm}\Gamma_2$ \cite{Vakhtel2023}. When the three leads are equally coupled (right panel), the landscape reveals a discrete set of extrema and saddle points. In the regime $\Gamma_3 \gg \Gamma_{1,2}$, the phase dependence is weaker and there is a minimum in $E_{01}$ at $\phi_{1,2} \approx \pi$. On the top of Fig. \ref{fig:PhaseBiasedTrijunction}, the Andreev energies $E_{0,1}$ corresponding to the trajectories in the $(\phi_1,\phi_2)$ plane indicated in the bottom panels are shown in dash-dotted (solid lines) for $\xi=0$ ($\xi=0.5$), that is, in (out of) resonance with the dot level. These energies are plotted with a $-\xi$ shift. We note that in the tunnel (low transparency) limit $\xi\gg\Gamma_\nu$'s we recover the conventional sinusoidal ground state dispersion $E_0(\{\phi_{\nu}\}) \approx -\big| \sum_{\nu} \Gamma_{\nu} e^{-i\phi_{\nu}}\big|^2/2\xi$, which reads $-(\Gamma_1\Gamma_2/\xi) \cos \phi_{12}$ in the two terminal case.

This two-level system would be an instance of an Andreev level qubit \cite{Desposito2001,Zazunov2003}, a kind of qubit defined with two Andreev states of a junction \cite{Janvier2015,Tosi2019,Hays2020,Hays2021}, and the choice of its operating point should consider aspects such as its sensitivity to noise. The noise in some parameter $\lambda$ produces decoherence through two mechanisms, pure dephasing and depolarization \cite{Ithier2005}. On the one hand, pure dephasing is related with the fluctuations in the qubit energy, blurring the access to the phase of superpositions of the type $c_0\ket{0}+c_1 e^{i\varphi}\ket{1}$, $c_{0,1}{\in}\mathbb{R}$. For small noise amplitudes, the associated rate $T^{-1}_{\varphi}(\lambda)$ is proportional to $|\partial E_{01} / \partial \lambda |^2$. On the other hand, the depolarization is mainly given by the relaxation from $\ket{1}$ to $\ket{0}$, and at small amplitudes the associated rate $T^{-1}_1(\lambda)$ is proportional to $|\bra{0}\partial H / \partial \lambda \ket{1}|^2$. Several strategies aim at reducing the effect of the noise: reducing its amplitude, correcting errors, or designing the qubit such that it acquires an inherent protection \cite{gyenis2021}. The latter approach can be explored when introducing different hardware configurations, attempting to reduce the dephasing and the depolarization rates simultaneously. This is not a  straightforward task and it must be noticed that simultaneous protection against noise in all parameters is impossible. In conventional qubits the noise in the charge offset and the external flux are emphasized as these are tunable parameters, but the noise in the Josephson coupling can also be important \cite{VanHarlingen2004,Ithier2005,Koch2007}. In the circuits we consider here, the couplings can be controlled by gate voltages, so the associated noise may be relevant too. For simplicity and to illustrate the principle, throughout this work we focus on the noise in the external flux, in the energy of the dot level and in the charge offset of the superconducting islands introduced in Sec. \ref{sec:island}. We provide further analysis of decoherence in Sec. \ref{sec:overview}. 

\subsection{Noise protection of the Andreev qubit}

The general challenge for the simultaneous reduction of dephasing and depolarization produced by noise in one parameter can be illustrated in the phase biased configuration of the present section. For example, the noise in $\xi$ has a dephasing sweet spot at $\xi=0$, where $E_{01}(\xi)$ is locally quadratic. However, at this spot the hybridization between the bare central level states $\ket{0}_d$ and $\ket{\uparrow\downarrow}_d$ is enhanced, so the relaxation rate increases (note that $\partial H^{\textrm{eff}}/\partial \xi \propto \tau_z$ and the states are $\propto \ket{0}_d\pm\ket{\uparrow\downarrow}_d$). The same occurs with the noise in the external flux, which has dephasing sweet spots where two states that were degenerate become hybridized. In the two terminal case, it occurs in the configuration of perfect transmission ($\xi{=}0$ and $\Gamma_1{=}\Gamma_2$ \cite{Vakhtel2023}, see top panels of Fig. \ref{fig:PhaseBiasedTrijunction}) at phase difference $\phi_e=\pi$, where the two degenerate states carry maximal and opposite supercurrents. There, a deviation in the energy level position or in the coupling symmetry hybridizes these states while flattening the transition energy at the same time. As will be discussed next, this analysis on the effect of flux noise must be extended to take into account the effect of the phase drop distribution.  

\begin{figure}[t]
    \centering
    \includegraphics[width=1\linewidth]{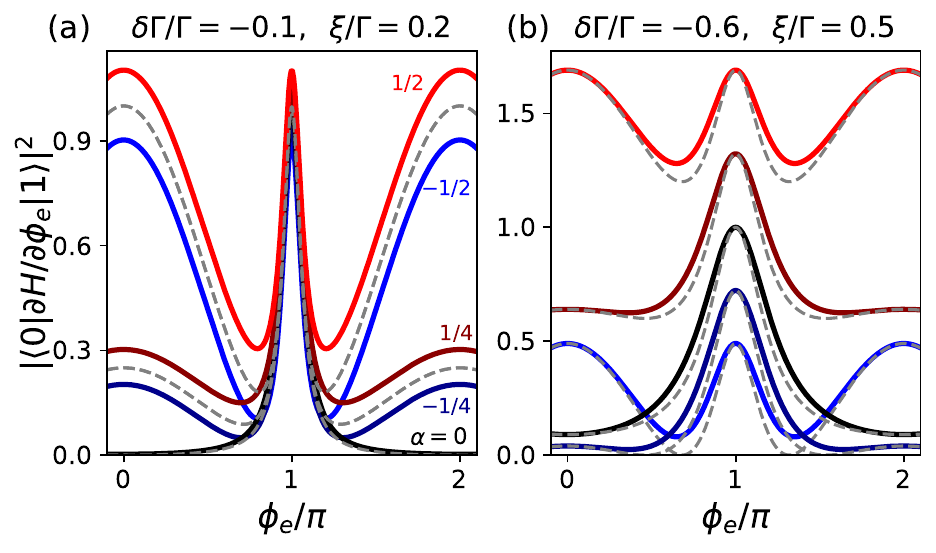}
    \caption{Phase dependence of the transition matrix elements (in units of $\Gamma{=}\Gamma_1{+}\Gamma_2$) for different phase drop distributions in a junction with two leads: symmetrically over the couplings ($\alpha{=}0$), on $\Gamma_1$ ($\alpha{=}-1/2$), on $\Gamma_2$ ($\alpha{=}1/2$) and halfway ($\alpha{=}\mp 1/4$). (a) Situation close to resonance. Low but finite deviation from $\xi{\neq}0$ and $\Gamma_1{\neq}\Gamma_2$ ($\delta\Gamma{=}\Gamma_1{-}\Gamma_2$) produces the high peak at $\delta{=}\pi$ due to the anticrossing between $E_0$ and $E_1$ (see Fig. \ref{fig:PhaseBiasedTrijunction} top). The slight asymmetry in the couplings produces asymmetry over the sign of $\alpha$ (gray dashed lines show the results for $\Gamma_{1,2}{=}1$). (b) Situation with larger coupling asymmetry, producing a large asymmetry over the sign of $\alpha$; dashed lines show the results for $\xi{=}0$.\label{fig:Gauge2T}}
\end{figure}

In a static situation, the phases in Eq. (\ref{Heffeven}) have a gauge freedom that manifests through the transformation $U {=} e^{if\tau_z/2}$, where the phase $f$ redistributes the phase drop over the couplings. In fact, note that the coupling term of $H^{\textrm{eff}}_{\textrm{even}}$ in Eq. (\ref{Heffeven}) can be written as $\sum_\nu \Gamma_\nu e^{i\phi_\nu \tau_z/2} \tau_x e^{-i\phi_\nu \tau_z/2}$, so the transformed coupling term in $UH^{\textrm{eff}}_{\textrm{even}}U^\dagger$ corresponds to a shift $\phi_\nu \rightarrow \phi_\nu+f$. However, if the external fluxes depend on time, such as in a situation with noise, the transformation adds to the new Hamiltonian the term $-i\hbar U (U^{-1})' = - \hbar f' \tau_z/2$, where the prime indicates time derivative and $f$ might depend on the internal and external parameters of the circuit. This non-equivalence means that a Hamiltonian with no term $\propto f'\tau_z$ would only be correct at a certain \textit{gauge choice} \cite{You-Sauls-Koch2019,Riwar2022}. Classically, the magnetic vector potential $\vec{A}$ --which determines the quantum phase drop through a Peierls substitution-- may be shifted by the gradient of a scalar function $F$ without altering the magnetic field, but if $F$ is time dependent, the complete gauge transformation must subtract $\partial_t F$ from the scalar potential $V$ --which determines a voltage profile-- so that the electric field $\vec{E}= -\vec{\nabla} V -\partial_t \vec{A}$ is not modified.  

The simplest way to notice this effect is in the flux noise matrix element of a junction with two leads. In this case, the general form of the coupling term in $H^{\rm eff}_{\rm even}$ is
 \begin{align*}
 &(\Gamma_1\cos{(-\phi_e/2{+}f)}+\Gamma_2\cos{(\phi_e/2{+}f)})\tau_x \nonumber \\ 
 & - (\Gamma_1\sin{(-\phi_e/2{+}f)} +\Gamma_2\sin{(\phi_e/2{+}f)})\tau_y,
%\label{eq:phase.biased.2}
 \end{align*}
where $f$ determines how asymmetrically this phase drops on the couplings between the leads and the dot. Supposing that this asymmetry is proportional to the phase drop, i.e. $f = \alpha \phi_e$, which would include terms $\propto \phi_e'\tau_z$ in the Hamiltonian for a general gauge, the squared matrix element $|\bra{0} \partial H / \partial \phi_e \ket{1}|^2$ is
\begin{align*}
\frac{(\Gamma_1^2{+}\Gamma_2^2{+}\xi^2)^2{-}4\Gamma_1^2\Gamma_2^2}{E_{01}^2} {-} \frac{\xi^2}{4} {+} \alpha(\Gamma_2^2{-}\Gamma_1^2) {+} \alpha^2 \left( \frac{E_{01}^2}{4}{-}\xi^2 \right),
\end{align*}
which is illustrated in Fig. \ref{fig:Gauge2T}. In panel (a), the couplings are almost symmetric and the matrix element depends slightly on the sign of $\alpha$ (the gray dashed lines correspond to the limit of $\Gamma_1{=}\Gamma_2$). In panel (b), the couplings are less symmetric and the sign of $\alpha$ has a strong effect reducing or boosting the relaxation (the gray dashed lines correspond to the limit of $\xi{=}0$). The matrix element at $\phi_e{=}\pi$ vanishes at $\alpha{=}\alpha_0 {:=} \Gamma/2\delta\Gamma$, a regime approached in the fully asymmetric phase-drop distribution $\alpha {=} {\pm}1/2$ at large coupling asymmetry. However, the value $\abs{\alpha_0} {\ge} 1/2$ requires that the phase drop winds more than once, an unlikely situation in a superconducting loop where vortices would appear to accommodate the next flux quantum whenever the phase surpasses $2\pi$ \cite{DeaverFairbank1961,DollNabauer1961,Kenawy2017}. The value at $\phi_e{=}\pi$ becomes a local minimum at $\alpha {=} \alpha_m$, with $2\alpha_m^2 {:=}(\xi^2{+}\Gamma^2)/(\xi^2{+}\delta\Gamma^2)$, with $\abs{\alpha_m} \geq 1/2$ too.

The particular $f$ value in each configuration is related with the electromagnetic field distribution in the circuit \cite{You-Sauls-Koch2019,Riwar2022,Bryon2023}, a calculation that is out of the scope of the present work. Our aim here is to identify the main features that determine it: the shape of the field in the central part of the junction depends on the geometry and properties of the circuit in which it is immersed \cite{Kenawy2022}, though in the situation we consider here we expect little variation in the field profile as the central region is small compared to a typical field wavelength and presumably homogeneous, i.e. we expect a symmetric drop of the phase difference. However, the state of the junction might carry a supercurrent, which modifies the field. A calculation in the lines of Refs. \cite{Zazunov2005,Kurilovich2021} (see Appendix \ref{ap:gauge}), which imposes electroneutrality in the electrodes, produces a deviation from the symmetric drop with $f {=} -\sum_\nu \Gamma_\nu \phi_\nu /\Delta$, where $\Delta$ is the superconducting gap. In the two terminals setup, it gives $\alpha=(\Gamma_1-\Gamma_2)/\Delta$, which vanishes in the limit of large gap.

In the setup with three leads [Fig. \ref{fig:sketch}d], the coupling term (Eq. (\ref{Heffeven})) is
 \begin{align*}
 &\Big[\Gamma_1\cos{(-\phi_{e1}/2{+}f)} + \Gamma_2\cos{(\phi_{e2}/2{+}f)} + \\ 
 & \Gamma_3\cos{(\phi_{e1}/2{-}\phi_{e2}/2{+}f)}\Big]\tau_x - \Big[\cos {\rightarrow} \sin\Big]\tau_y,
 \end{align*}
\noindent
which in the limit $\Gamma_3 {\rightarrow} 0$ is not equivalent to the two leads case: if the fluxes have the same source and obey $\phi_{1,2e} = \eta_{1,2} \phi_e$, then $f = (\Gamma_1\eta_1-\Gamma_2\eta_2)\phi_e/2\Delta$. For $\eta_{1,2}=1$, we have the same Hamiltonian as in the two leads case, but now the phase drops $-\phi_e/2{+}(\Gamma_1{-}\Gamma_2)\phi_e/2\Delta$ and $\phi_e/2+(\Gamma_1-\Gamma_2)\phi_e/2\Delta$ can go up to $2\pi$, therefore becoming able of achieving the relaxation free point $\alpha_0$ for parameters where $1/2\leq \abs{\alpha_0} \leq 1$.

We note that there are other transformations $U_n {=} e^{if_n\tau_n}$ ($n{\neq}z)$ that generate the time dependent terms $-\hbar f_n' \tau_n/2$. The case $n{=}0$ ($\tau_0{=}\mathbbm{1}$) refers to a global energy shift and introduces a global time dependent phase with no effects. The cases $n{=}x,y$ have no apparent physical interpretation, but the time dependent term must be kept if the transformation has been made for example to facilitate calculations. 

As argued in the previous paragraphs, the phase drop distribution is very important for the relaxation rates. In addition, from a broader perspective, its effect is pervasive on other observable quantities that depend on the matrix elements, such as the shift of a resonator coupled to the circuit or the transition rates between the states in the junction \cite{Olivares2014,Park2017,Trif2018,Park2020,Metzger2021,Hermansen2022,Ackermann2023,Fauvel2023}.

\subsection{Topological properties}

This model of the trijunction in a perfect phase-biasing configuration can be linked with the topological singularities predicted in general multiterminal junctions \cite{Riwar2016}, since there are certain configurations where $E_{01}$ reaches $0$, forming a Weyl point. In our case (see Eq. (\ref{spectrum})), these points appear when the parameter $\xi$ and the sum $\Gamma_T = \sum_\nu \Gamma_\nu e^{i\phi_\nu}$ vanish, so topological protection in this model against variations in $\xi$ is not possible, though it is on the $\Gamma_\nu$'s and the $\phi_\nu$'s. It is convenient to interpret $\Gamma_T$ as a vector in the complex plane, formed by the addition of the constituent numbers of magnitude $\Gamma_\nu$ and direction $e^{i\phi_\nu}$. In the case of two terminals, the only possibility is $\Gamma_1=\Gamma_2$ with a phase difference of $\pi$. However, with three terminals, if the $\Gamma_\nu$'s satisfy the triangle inequality (each side is smaller than the sum of all the other sides) there is a set of phase differences that produce a Weyl point -- a similar argument has been recently used to map a general model of a multiterminal quantum dot Josephson junction into a two terminal one \cite{Zalom2023}. In this case, the presence of Weyl points would be protected against variations on the $\Gamma_\nu$'s, as long as they continue satisfying the triangular equality. If the number of terminals is increased to $\mathcal{N}$, the generation of Weyl points becomes more probable since there are more ways of closing a polygon with $\mathcal{N}$ sides, and they live in a phase space of $\mathcal{N}-1$ dimensions. 

This configuration can be compared to a multiterminal junction in the tunnel limit, with $H = \Re{\sum_{\mu>\nu} E_{J_{\mu\nu}} e^{i(\phi_\mu-\phi_\nu)}}$. In this case there is only one ``band'' but the points where $H=0$ are interesting as they correspond to a maximum supercurrent, and the $E_{J_{\mu\nu}}$'s play a role similar to that of the $\Gamma_\nu$'s for defining these points.

\section{Charge islands' configuration}
\label{sec:island}

In this section we consider the situation in which the central region is connected to isolated superconducting islands [Fig. \ref{fig:sketch}e]. In this configuration the number of Cooper pairs $N_\nu$ in each island $\nu$ is a discrete variable and the total number of pairs $N_T = \sum_\nu N_\nu + n_d$ is conserved, where $2n_d$ is the number electrons in the central region. This conservation constrains the accessible states in the basis $\ket{\{N_\nu\},n^m_d}$ for a given value of $N_T$, where the $\ket{n^m_d}$ are the states in the central region. In our model, we consider only $\ket{0}_d$ ($n_d{=}0$) and $\ket{\uparrow\downarrow}_d$ ($n_d{=}1$), with $\hat{n}_d = \sum_\sigma d^\dagger_\sigma d_\sigma/2$, so the label $m$ will be dropped from now on. As a result of the constraint, an instance (value) of $N_T$ can be described with a smaller basis, e.g., one lead variable can be removed using $\hat{N}_\mu=N_T-\sum_{\nu\neq\mu}\hat{N}_\nu-\hat{n}_d$.

The charging effects in and between the islands are described by a capacitance matrix $\mathcal{C}$ such that the associated Hamiltonian is $H_C = (\vec{N}-\vec{n}_g)^T \mathcal{C}^{-1}(\vec{N}-\vec{n}_g)$, where $\vec{N} = (\hat{N}_1,\hat{N}_2,...,\hat{N}_\mathcal{N})$ is the vector of the number operators in each of the $\mathcal{N}$ islands and $\vec{n}_g$ is composed of the corresponding effective charge offsets, which indicate the charge configuration in a situation of electrostatic equilibrium. In general, $H_C$ can be interpreted as a confining potential in the charge basis, and it could also include capacitive couplings between the leads and the central region by increasing its size and including $\hat{n}_d$ in $\vec{N}$. The particular values of $\mathcal{C}$ depend on how the electromagnetic field accommodates into the geometry and materials of the device. We consider the regime where the absolute charge offsets are large enough so that the eigenspectrum of the $\hat{N}_\nu$'s can be extended to $N_\nu \in \mathbb{Z}$ for convenience (the $n_{g_\nu}$'s are shifted to denote deviations from the integer set closest to equilibrium).

\begin{figure}[t]
    \centering
    \includegraphics[width=1\linewidth]{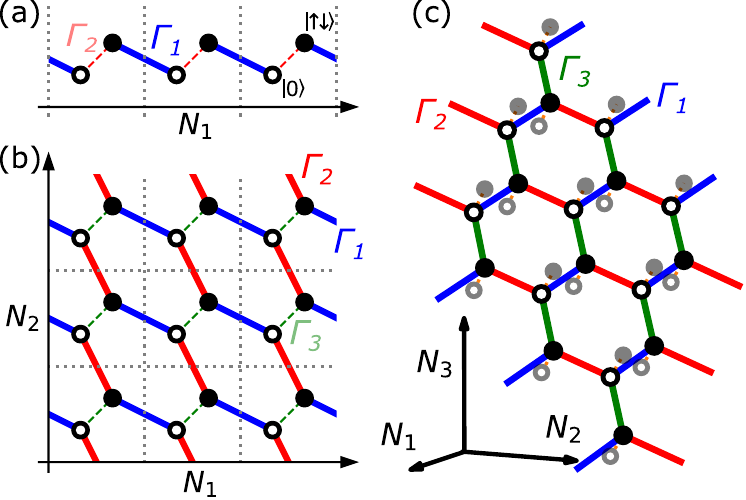}
    \caption{Lattice representation of the island configuration. The island charge variables $N_\nu$ define the dimensions of the {\it crystal} and the fermionic states are the atoms in the unit cell (dotted gray lines), with hollow (full) circles for $\ket{0}$ ($\ket{\uparrow\downarrow}$). The sets of connected states for an instance (value) of total charge $N_T$ are the blue dimers in (a), the blue-red chains in (b) and the blue-red-green grid in (c) for the setups of 1, 2 and 3 leads, respectively. The dashed lines represent the projection of a single instance in a space with lower dimension, where the projected variable is determined by the conservation of $N_T$, {\it e.g.} a blue-red chain in (a), where $N_2 = N_T-N_1-n_d$. The pale circles in (c) represent sites of adjacent $N_T$ grids, which would be connected in the presence of a fourth lead. \label{fig:number_sketch}}
\end{figure}

As discussed in Ref. \cite{Vakhtel2023}, deep inside the gap and restricted to the even parity sector, the main tunneling process is the effective interchange of Cooper pairs between the leads and the central level, though microscopically the electrons actually tunnel one by one  \cite{Keselman2019}. Thus, in the infinite gap limit discussed in the previous section, the coupling Hamiltonian is $H_c = \sum_\nu \Gamma_\nu e^{-i \hat{\phi}_\nu} d^\dagger_\uparrow d^\dagger_\downarrow + h.c.$, where $e^{-i\hat{\phi}_\nu}$ decreases $N_\nu$ by $1$ (now the phase is the generator of the charge translations with $[ \hat{\phi}_\mu , \hat{N}_\nu]=i\delta_{\mu\nu}$), and $d^\dagger_\uparrow d^\dagger_\downarrow$ introduces both electrons of the pair in the dot. The $N_T$ constraint in the Hilbert space $\ket{\{N_\nu\},n_d}$, which can be described by a gauge symmetry in the phase space \cite{Vakhtel2023,Karki2023}, may be noticed by the fact that the states $\ket{\{N_\nu\},0}$ and $\ket{\{N_\nu\},\uparrow\downarrow}$ are not connected in any way, because the coupling conserves the total charge. However, these states are hybridized when an additional island is introduced -- this is the main feature explored in this section.

\begin{figure*}[t]
    \centering
    \includegraphics[width=2\columnwidth]{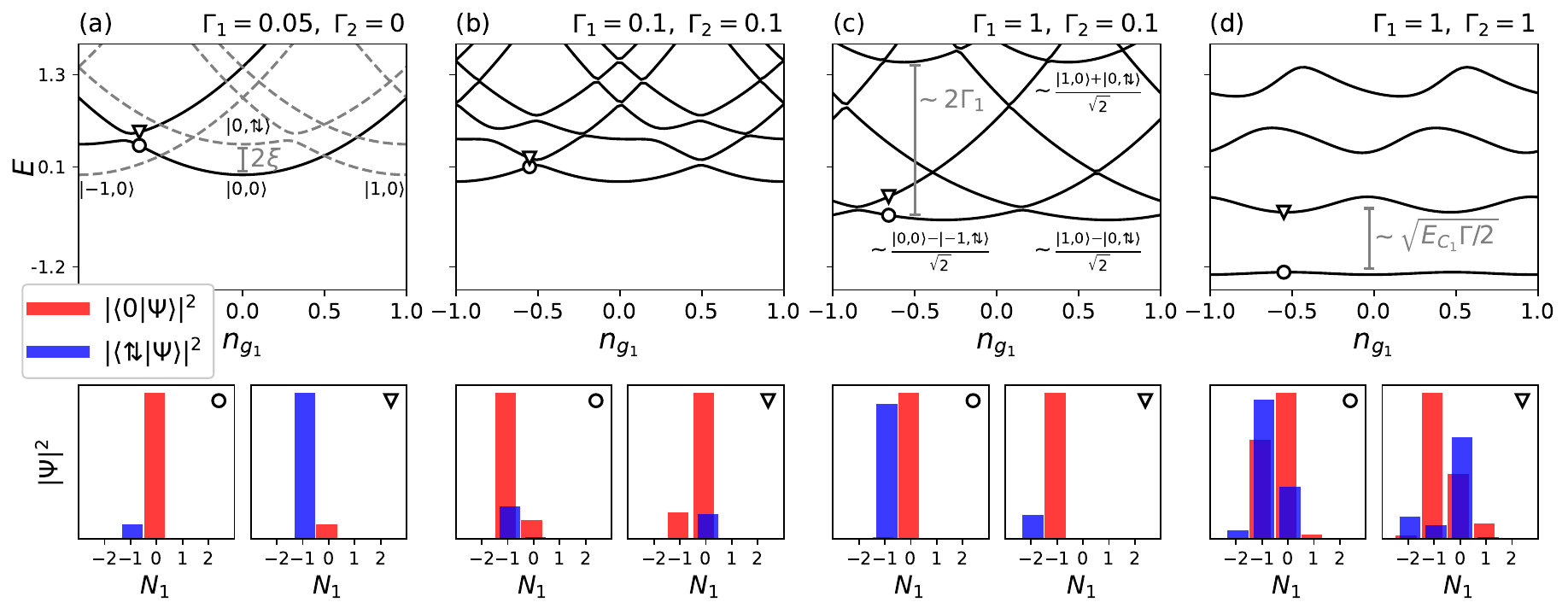}
    \caption{Energy dispersion over the charge offset $n_{g_1}$ in the configurations with 1 and 2 terminals ($E_{C_2}=0$, $\xi=0.2$, energies in units of $E_{C_1}$), and wavefunctions indicated with markers (bottom panels). (a) Small coupling of the central level with island $1$: the instance $N_T{=}0$ is in solid lines, and other instances in dashed; some bare $\ket{N_1,n_d}$ states are indicated close to the bottom of their charging parabolas. (b) Small couplings with islands $1$ and $2$: finite $\Gamma_1$ and $\Gamma_2$ connect all states in (a). (c) Large (small) coupling with terminal $1$ ($2$): $\Gamma_1$ dominates the formation of symmetric and antisymmetric combinations. (d) Large couplings (more similar to a transmon except that the junction is close to resonance, not in the tunnel regime). %(Josephson-Andreev transmon -- {\it JAmon}).
    \label{fig:1and2T}}
\end{figure*}

If the basis is portrayed as a lattice in the $\{N_\nu\}$ space with a unit cell comprising the fermionic levels, each $N_T\in\mathbb{Z}$ constitutes an independent set disjoint from the others (see Fig. \ref{fig:number_sketch}). Starting with one island and the central level, these sets are the groups of two sites fulfilling $N_1+n_d=N_T$ (the dimers connected with blue lines in Fig. \ref{fig:number_sketch}a). Then, when the second island is coupled, the sets are 1D {\it chains} defined by $N_1+N_2+n_d=N_T$ (the blue and red chains in Fig. \ref{fig:number_sketch}b). One of these instances of $N_T$ is projected on panel (a), where the dashed red lines represent the coupling $\Gamma_2$ and $N_2=N_T-N_1-n_d$. When a third lead is connected, the chains connect between themselves through $\Gamma_3$ and form a 2D {\it grid} fulfilling $N_1+N_2+N_3+n_d=N_T$, depicted in Fig. \ref{fig:number_sketch}c and projected to $N_{1,2}$ in panel (b). Similarly, these grids may be connected when a fourth lead is included, and so on. In the following subsections, this kind of circuits is progressively explored along with some applications for the definition of protected qubits.

\subsection{One terminal}
\label{sec:island.1}

We consider first a single superconducting island with $H_C=E_{C_1}(\hat{N}_1{-}n_{g_1})^2$ connected to the central region. The relevant states $\ket{N_1,n_d}$ are 
\begin{align}
    \ket{0,0} \rightarrow E&=E_{C_1}n_{g_1}^2  \nonumber \\
    \ket{-1,\uparrow\downarrow} \rightarrow E&= E_{C_1}(-1-n_{g_1})^2 + 2\xi,
\end{align}

\noindent
where a translation in $N_1{=}N_T$ would refer to other instances of total charge (blue dimers in Fig. \ref{fig:number_sketch}a). The energies of the $N_T{=}0$ instance ($n_{g_1}$ is shifted so that it refers to the instance with lowest energy) are the solid black parabolas in Fig. \ref{fig:1and2T}a, hybridized by the tunneling $\Gamma_1$, which opens an anticrossing at $n_{g_1} {=} -1/2-\xi/E_{C_1}$ (close to the symbols that indicate the wavefunctions in the lower panels). The dashed lines in Fig. \ref{fig:1and2T}a represent other instances of $N_T$, which are in principle uncoupled subspaces.

As indicated in the lower panels, the two-level system in the limit $\Gamma_1 \ll E_{C}$ is similar to a charge qubit with a steep $n_{g}$ dependence that would be detrimental for coherence in the presence of charge noise. In the opposite limit $\Gamma_1 \gg E_{C}$, the eigenstates approximate the symmetric and antisymmetric combinations $\ket{N_1',\pm} {=} (\ket{N_1,0}{\pm}\ket{N_1{-}1,\uparrow\downarrow})/\sqrt{2}$ for a larger range in $n_{g_1}$, with a transition energy $\sim2\Gamma_1$. The dispersion of $E_{01}$ over $n_{g_1}$ is reduced (${\sim}E_{C_1}^2n_{g_1}^2/\Gamma_1$, disregarding $\xi$), but now the relaxation produced by the noise in $n_{g_1}$ is larger as $\partial H / \partial n_{g_1}$ is non-diagonal in the qubit basis. When compared with the transmon, this setup does not suffer the problem of reduced anharmonicity, but we note that the single level model of Eq. (\ref{Heff}) would not be valid at arbitrary $\Gamma_1$ because the qubit states could approach other levels in the central region. A more microscopic model of this configuration has been studied in \cite{Pavesic2021}. 

\subsection{Two terminals}
\label{sec:island.2}
When a second terminal is connected to the central region, the constraint $N_1+N_2+n_d=N_T$ produces independent sets with an infinite number of states, the blue and red chains in Fig. \ref{fig:number_sketch}b. In the extended basis $\ket{N_1,N_2,n_d}$, the coupling $\Gamma_2$ (red) connects $\ket{0,0,0}$ with $\ket{0,-1,\uparrow\downarrow}$, which in the projected basis $\ket{N_1,n_d}$ appears as a coupling between $\ket{0,0}$ and $\ket{0,\uparrow\downarrow}$ (red dashed lines in Fig. \ref{fig:number_sketch}a). To illustrate this we start by considering that the second terminal is very large so that there is no additional charging energy. 

In this situation, we have $H_C=E_{C_1}(\hat{N}_1{-}n_{g_1})^2$, and in the limit $\Gamma_{1,2} \ll E_{C_1}$, the states (in the reduced basis) $\ket{0,0}$ and $\ket{-1,\uparrow\downarrow}$ maintain the anticrossing given by $2\Gamma_1$ as in the previous subsection. The coupling $\Gamma_2$ does not directly open anticrossings between $\ket{N_1,0}$ and $\ket{N_1,\uparrow\downarrow}$, because their energies never coincide. However, at $n_{g_1}{=}{-1/2}$ (indicated by symbols in Fig. \ref{fig:1and2T}b) both couplings combine to hybridize $\ket{N_1{-}1,n_d}$ with $\ket{N_1,n_d}$ through an intermediate state with opposite central level occupation. For finite $\abs{\xi} \gg \Gamma_{1,2}$ the gap opening is $\sim (\Gamma_1^2+\Gamma_2^2)/\xi$. The states $\ket{N_1{-}1,0}$ and $\ket{N_1,\uparrow\downarrow}$ also hybridize through two intermediate states at $n_{g_1} {\approx} -1/2+\xi/E_{C_1}$, with an even smaller anticrossing (the size decreases with the number of intermediate states needed for the indirect coupling). In the lower panels we show the different states $|N_1\rangle$ contributing to the ground and the first excited eigenstates. As we discuss next, more and more charge states contribute as the coupling is increased.

\begin{figure*}[t]
    \centering
    \includegraphics[width=2\columnwidth]{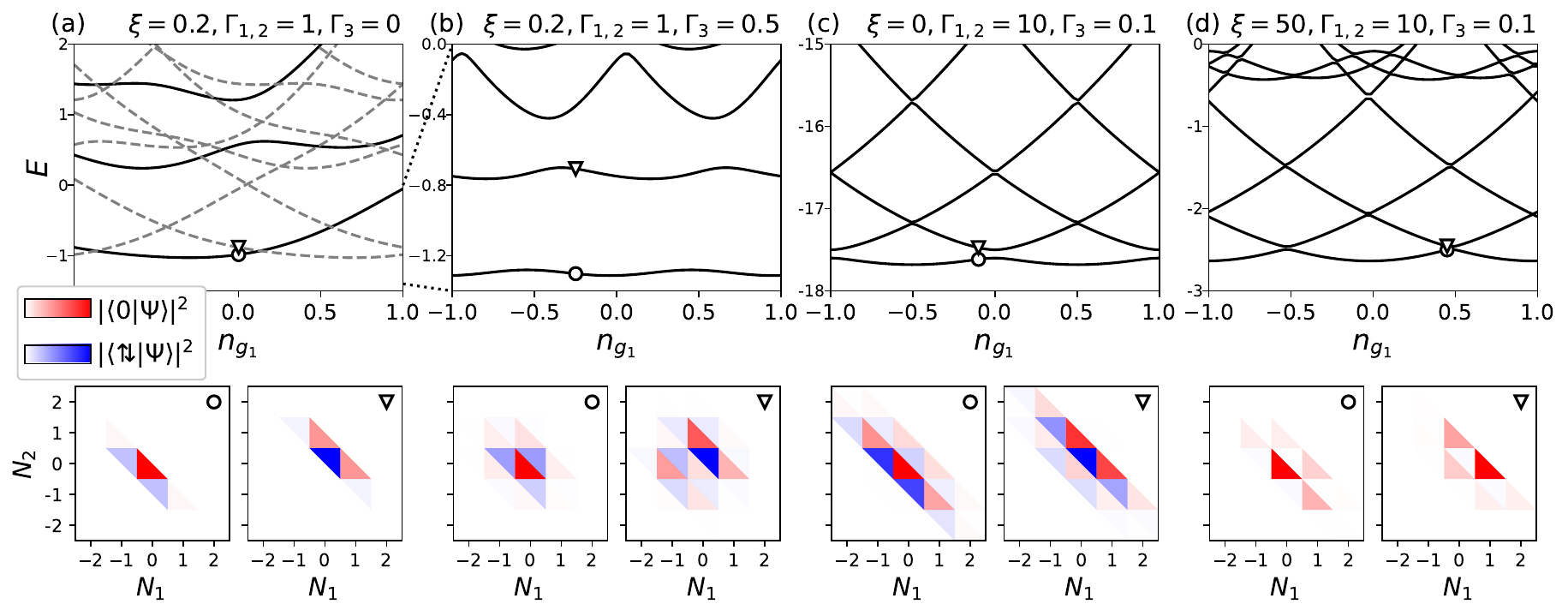}
    \caption{Energy dispersion over the charge offset $n_{g_1}$ in the configurations with two and three terminals when two of them are islands with finite charging energy ($n_{g_2}=0$, $E_{C_3}{=}0$, energies in units of $E_{C_1}=E_{C_2}$) and wavefunctions indicated with markers, represented in a $N_1$-$N_2$ space where each cell is divided into a lower and an upper triangle for the level states $\ket{0}$ and $\ket{\uparrow\downarrow}$, respectively (bottom panels) (a) Only two terminals as in Fig. \ref{fig:1and2T}d but now terminal $2$ has finite charging energy, which renders the spectrum aperiodic and non-equivalent between instances ($N_T{=}0$ in solid lines, others in dashed). (b) Introducing the coupling $\Gamma_3$ with the third terminal, which connects the different instances (chains) in (a). (c) Large couplings $\Gamma_{1,2}$ (transmon regime in the $N_1{-}N_2$ direction) and small coupling $\Gamma_3$ (Cooper pair box regime in the orthogonal direction), central level close to resonance. (d) Same with central level far from resonance (tunneling regime).
    \label{fig:3T}}
\end{figure*}

In the limit $\Gamma_1 {\gg} E_{C_1} {\gg} \Gamma_2$ (corresponding qualitatively to Fig. \ref{fig:1and2T}c), the lowest energy states, ${\sim} \ket{N_1',-}$ and ${\sim} \ket{N_1'{+}1,-}$, anticross close to $n_{g_1}{=}N_1$ opening a gap ${\sim} \Gamma_2$. Note that these two levels are completely disjoint at $\Gamma_2{=}0$, defining a qubit protected against relaxation by charge noise. However, as discussed in the previous subsection, the decoherence produced by the noise in $n_{g_1}$ may be important for feasible values of $\Gamma_1$.

We address now the regime $\Gamma_{1,2} {\gg} E_{C_1}$ in which the wavefunctions extend broadly over the charge states [Fig. \ref{fig:1and2T}d], thereby decreasing the modulation produced by the offset charge $n_{g_1}$, just like in the transmon \cite{Koch2007}. We note that transmon implementations in hybrid semiconductor-superconductor nanostructures provide gate tunability. The resulting \textit{gatemons} \cite{Larsen2015,deLange2015,Aguado2020} are thus more suitable to manifest features from the Andreev structure, in particular when tuned close to resonance \cite{Bargerbos2020,Kringhoj2020}. The Josephson-Andreev junction should recover the typical tunnel junction transmon when the central level is far from resonance. In this case there is an exponential suppression of the charge dispersion with $E_J^{\textrm{eff}}/E_{C_1}$, where $E_J^{\textrm{eff}}$ is the effective amplitude of Cooper pair tunneling between the leads \cite{Koch2007}. In the resonant limit, the Josephson-Andreev transmon (\textit{JAmon}) dispersion is even more suppressed and becomes dominated by the tunneling of pairs of Cooper pairs \cite{Vakhtel2023,Bargerbos2020,Kringhoj2020}. In either case, the dispersion is a non-perturbative effect. 

In order to estimate the levels energies, we may disregard this dispersion and use an adiabatic approximation in the phase space, which consists in using the phase dependence $E_0(\phi_e)$ (Eq. (\ref{spectrum})) of the Andreev lowest state as the effective potential in the phase variable $\phi_e \rightarrow \hat{\phi}_1$. Then, in the transmon-like limit of localized phase, it is expanded up to fourth order at $\phi_e{=}0$ to get the harmonic separation $\omega \approx E_{01}$ between states and the anharmonicity $\alpha_h {=} E_{12}{-}E_{01}$ at lowest order. In the tunnel limit $\xi {\gg} \Gamma_{1,2}$, $\omega \approx 2\sqrt{E_{C_1}\Gamma_1\Gamma_2/2\xi}$, and in the resonant limit $\delta\Gamma{=}0{=}\xi$, with $\Gamma, \delta\Gamma{=}\Gamma_1{\pm}\Gamma_2$, $\omega \approx 2\sqrt{E_{C_1}\Gamma/8}$. More details are provided in Appendix \ref{ap:adiabatic}.  

\subsubsection*{Two charge islands}
In the situation we have analized so far, where there is charging energy only in one island, the spectrum is periodic in $n_{g_1}$, as a charge translation in $N_1$ equals to a charge translation in the charge offset. Moreover, all instances $N_T$ are equivalent, because a charge translation in $N_1$ also corresponds to a translation in $N_T$. This also occurs in the situation where the charging energy only depends on the imbalance between the two islands, with $H_C=E_{C_{12}}(\hat{N}_1{-}\hat{N}_2{-}n_{g_{12}})^2$, which has been thoroughly analyzed in Ref. \cite{Vakhtel2023}. However, this is not true in general when the two leads are islands with finite charging energies. 

In this case, the additional charging term in the reduced basis,
\begin{equation*}
E_{C_2}(\hat{N}_2{-}n_{g_2})^2 = E_{C_2}(N_T{-}\hat{N}_1{-}\hat{n}_d{-}n_{g_2})^2,  
%\label{eq:EC2}
\end{equation*} 
removes the periodicity in the offset charges and the equivalence between $N_T$ instances. This is shown in Fig. \ref{fig:3T}a, which uses the same parameters as in Fig. \ref{fig:1and2T}d except for the introduction of $E_{C_2}=E_{C_1}$. We observe in the lower panels of the figure that now the wavefunctions corresponding to two different instances delocalize along lines of fixed total charge in the ($N_1$-$N_2$) plane (see Fig. \ref{fig:number_sketch}b). The transition to the three leads case (see next subsection) is similar to the previous discussion when going from one to two leads in Fig. \ref{fig:1and2T}a, where different instances become coupled by the additional lead.

\subsection{Three terminals}

When a third terminal is connected to the central region, there are two independent charge variables -- we may use for example the projected basis $\ket{N_1,N_2,n_d}$ where $\hat{N}_3=N_T{-}\hat{N}_1{-}\hat{N}_2{-}\hat{n}_d$. In this situation, if the charging potential is flat in one direction, the wavefunctions are not constrained and a continuous band is formed in that direction. Considering the charging terms as a quadratic form in the charge variables, then the previous condition occurs when one of its principal axes has a $0$ eigenvalue. Therefore, in order to have a discrete set of states, we consider a fully confining potential. In particular, we select the potential for the situation where two terminals are islands with finite charging energies $E_{C_1}$ and $E_{C_2}$ (used in the last part of the previous subsection), and the third terminal has no charging energy.

In this case, a finite coupling $\Gamma_3$ connects the chains that were independent instances in the case with two terminals [Fig. \ref{fig:3T}a,b], and the wavefunctions extend in the number space. When the wavefunctions are broadly extended in charge, the system can be thought as a set of coupled harmonic oscillators, a limit which is analyzed in the tunneling regime in Sec. \ref{sec:cotunneling}. Here, we focus in the limit of small $\Gamma_3$ compared with the charging energy in the transverse direction of the chains, which suggests the definition of a qubit formed by two adjacent chains [Fig. \ref{fig:3T}c,d]. This limit is better understood in a number basis where one degree of freedom refers to the extension of the wavefunction on the direction of a chain ($N_{12}:=N_1{-}N_2$, the diagonal in the wavefunction plots in Fig. \ref{fig:3T}) and the other one refers to the specific chain ($N_1{+}N_2{+}n_d = N_T{-}N_3$, the antidiagonal). The offset charge does not affect in the chain direction, but in the orthogonal one it produces charge parabolas. In the case $E_{C_2}{=}E_{C_1}$, illustrated in Fig. \ref{fig:3T}, the charging potential in that orthogonal direction is  $\sim E_{C_1}(N_3{+}n_d{+}N_T{+}n_{g_1}{+}n_{g_2})^2/2$. Due to the structure of the wavefunctions in the resonant and tunneling (large $\xi$) limits, where $\langle n_d \rangle \approx 1/2,\,0$, respectively [panels c and d], the spectrum has minima in $n_{g_1}$ at $\mathbb{Z}{+}1/2$ and $\mathbb{Z}$ when setting $n_{g_2}{=}0$. The structure of the wavefunctions also determines the size of the anticrossings, which is $\sim 2\Gamma_3$ in the resonant and $\sim \Gamma_3^2/\xi$ in the tunneling limit.

\subsection{Discussion}
\label{sec:islands.discussion}
The lattice picture used in this section of charge islands allows us to relate these systems with the definition of protected qubits that use states of disjoint support. As argued in a recent work that addresses the protection enabled by the resonant regime of a quantum-dot junction in a loop \cite{Vakhtel2023flux} (see Sec. \ref{sec:loop2}), the idea of this protection can be represented in a general manner by a 1D lattice where the nearest neighbour couplings vanish. In that situation, if only next-nearest neighbour couplings are present, the lattice disconnects into two separate chains, which become the hosts of the disjoint states that define the qubit. In our case, the lattice for the Josephson-Andreev junction with three leads, where one of them profits from a controllable coupling with the central level [Fig. \ref{fig:number_sketch}b,c], enjoys a similar interpretation. Here, two adjacent chains formed by the arrangement of two leads host the disjoint qubit states. The difference is that the coupling between the chains, mediated by the coupling $\Gamma_3$ with the third lead, is not completely equivalent to the reassembling of a single chain but skips alternate nearest neighbour couplings and may also involve other chains. Additionally, in this system the only parameter that controls the coupling between chains is $\Gamma_3$, a property that seems to be model independent because no matter the particular Andreev structure, the conservation of the total charge forbids interchain coupling if the third lead is disconnected. This applies to the introduction of more levels in the modeling of the central region, quasiparticle states in the leads, or direct tunneling between the leads (probably relevant for the implementation of Fig. \ref{fig:sketch}a). 

The problem with this kind of protected Hamiltonian structures is that the parameter region where the disjointness takes place is small: the resonant regime of the quantum-dot fluxonium requires vanishing $\xi$ and $\delta\Gamma$ \cite{Vakhtel2023flux}, and the multiterminal island configuration with two independent chains requires vanishing $\Gamma_3$ for the disconnection and fine tuning of the charge offsets for the degeneracy. These parameters are also a source of dephasing, because the transition energy varies linearly from the anticrossing. In particular, the charge noise from the charge offsets $n_{g_\nu}$'s is important because the chains are charge states like those in a Cooper pair box with a tunable Josephson coupling determined by $\Gamma_3$. 

Finally, as a curious property, we point out that our microscopic model of the junction bestows a \textit{polyatomic} unit cell to the number lattice. For instance, in the case with two leads the Hamiltonian equals a Su–Schrieffer–Heeger (Rice-Mele) model for $\xi{=}(\neq)0$ [Fig. \ref{fig:number_sketch}a] when there are no charging energies \cite{SSH79,RiceMele82}. Increasing the number of states describing the microscopic part of the junction would increase the number of sites in the unit cell, implementing different kinds of crystal Hamiltonians of $\mathcal{N}{-}1$ dimensions, where $\mathcal{N}$ is the number of leads. However, getting evidence of edge states in a topological phase would require the use of a charging potential able to provide a steep barrier in the number coordinate while maintaining a flat potential in the interior of the well.

\section{Loop configuration}
\label{sec:loop}

\begin{figure*}[t]
    \centering
    \includegraphics[width=\linewidth]{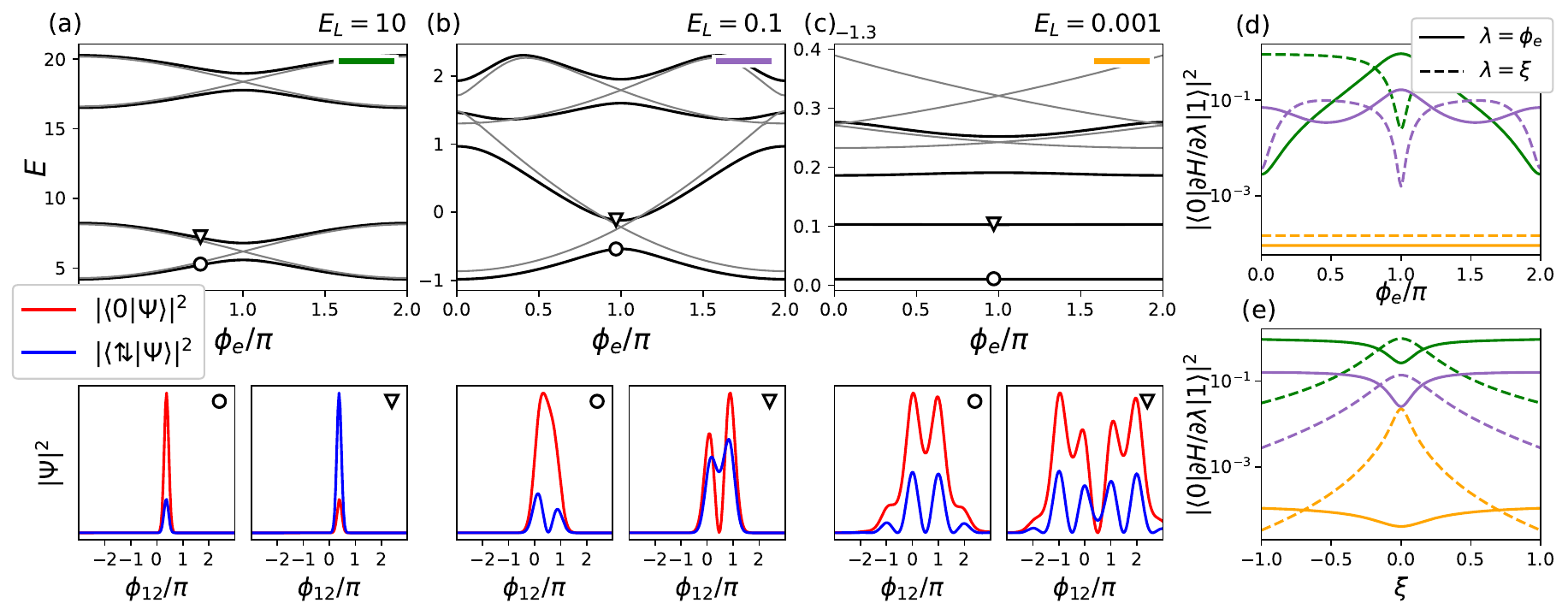}
    \caption{(a-c): Energy dispersion over the external phase $\phi_e$ for decreasing values of $E_L$ in the loop configuration with two leads, and wavefunctions indicated with markers (bottom panels -- they are centered over $\phi_{12}\sim \phi_e/2$ because of the form of the inductive term $E_L(2\phi_{12}-\phi_e)^2$). Grey lines indicate the resonant case with $\Gamma_{1,2}=1$ and $\xi=0$ (energies in units of $E_{C_{12}}$), and black lines indicate the case $\Gamma_{1}=0.95$, $\Gamma_{2}=1.05$ and $\xi=0.6$ (with a shift of $-\xi$). (d,e): Matrix elements associated to noise in $\phi_e$ and $\xi$ for the non-resonant parameters in a-c (setting $\phi_e=0.95\pi$ in e) and $\alpha{=}0$, indicating each case with the colours in the a-c corners. \label{fig:2fluxanium}}
\end{figure*}

We now consider the case where two leads are connected forming a loop [Fig. \ref{fig:sketch}f], for which a different, more macroscopic model is required. We use the variable of tunneled pairs between the leads of the loop $\hat{N}_{12} {=} \hat{N}_1{-}\hat{N}_2$, conjugate to $\hat{\phi}_{12}$, with charging energy $E_{C_{12}}\hat{N}_{12}^2$. This model without the separate islands 1 and 2 is qualitatively different from the one derived in the previous section, because the loop disables the island nature of these leads, rendering the charge difference continuous \cite{Devoret2021} and allowing the presence of high energy persistent current states \cite{Koch2009}. The charge in the remaining island $3$ remains discrete, and the total Hamiltonian is the combination $H_{12} + H_3 + H_d + H_c$:
\begin{align}
    H_{12} & = E_{C_{12}} \hat{N}_{12}^2 + E_L(2\hat{\phi}_{12}-\phi_e)^2 \nonumber \\
    H_3 & = E_{C_3} (\hat{N}_3 - n_{g_3})^2 \nonumber \\
    H_d &  = \epsilon \sum_{\sigma} d^\dagger_\sigma d^{}_\sigma + u d^\dagger_\uparrow d^{}_\uparrow d^\dagger_\downarrow d^{}_\downarrow \nonumber \\
    H_c &  = \left(\Gamma_1 e^{-i\hat{\phi}_{12}} + \Gamma_2 e^{i\hat{\phi}_{12}} + \Gamma_3 e^{-i\hat{\phi}_3}\right)d^\dagger_\uparrow d^\dagger_\downarrow + h.c.  \, ,
    \label{eq:HT_loop_and_island}
\end{align}
\noindent
where the Josephson processes that couple the dot with the leads in $H_c$ take into account that a tunneling of a pair from $\nu{=}1(2)$ to the dot level produces a shift in $N_1-N_2$ by $\mp1$ (translation $e^{\mp i\hat{\phi}_{12}}$) and that a tunneling from $\nu=3$ to the dot level produces a negative shift in $N_3$ (translation $e^{-i\hat{\phi}_3}$). The Hamiltonian can be solved in the $\ket{\phi_{12},N_3,n_d}$ basis with $\hat{N}_{12}=-i\partial_{\phi_{12}}$.

The phase drop $\phi_e = \Phi/\Phi_0$ is to be distributed among the inductance and the couplings $\Gamma_{1,2}$. In a dynamical situation, the correct description for a tunnel junction without 
Andreev structure corresponds to the phase dropping completely on the inductance \cite{You-Sauls-Koch2019, Bryon2023} (in the absence of additional time dependent terms). However, in our case, a freedom in the drop distribution within the junction is still present ($\Gamma_{1,2} \rightarrow e^{if}\Gamma_{1,2}$ in $H_c$, Eq. (\ref{eq:HT_loop_and_island})). We note that the inductive potential $E_L(2\hat{\phi}_{12} - \phi_e)^2$ includes a factor of $2$ as the complete transfer of a pair from lead $1$ to lead $2$ requires two tunneling events. In the limit $L {\rightarrow} 0$ ($E_L{\rightarrow}\infty$), the phase becomes localized at $\hat{\phi}_{12} {\rightarrow} \phi_e/2$, recovering the model in section \ref{sec:phase_bias}. 

From a circuit perspective, the loop-island Hamiltonian contains two modes, one flux-like and one charge-like, which may be harnessed to devise protected qubits. These qubits would be similar in nature to the ones conceived in multimode circuits where several conventional elements generate the different modes \cite{gyenis2021}. In the remaining of this section, we explore the full Josephson-Andreev Hamiltonian, first in a one loop configuration, and then introducing a third floating lead. In section \ref{sec:cotunneling}, we focus on the cotunneling limit, which allows a straight-forward correspondence with the protected multi-element circuits.

\subsection{Two terminals forming one loop}
\label{sec:loop2}

This configuration is similar to a flux qubit when the Josephson-Andreev junction is in the tunneling regime \cite{Orlando1999,vanderWal2000,Friedman2000,Manucharyan2009}. Otherwise, when the Andreev structure is close to resonance ($\Gamma_1{=}\Gamma_2$, $\xi{=}0$), it has been recently identified as a prospective protected qubit based on two disconnected states \cite{Caceres2022,Vakhtel2023flux}.

In Fig. \ref{fig:2fluxanium}, we review the main features of the energy spectrum of this system, along with the corresponding wavefunctions. When $E_L$ is large compared with $E_C$ [Fig. \ref{fig:2fluxanium}a], the phase is localized and we recover the Andreev states of the central level from Section \ref{sec:phase_bias}. We note that the fermionic structure already provides a difference with the flux qubits, as the first excited level is an Andreev state. The associated wavefunctions (lower panels) are hybridizations between $\ket{0}$ and $\ket{\uparrow\downarrow}$ at a localized phase. In contrast, the excited states of the flux qubit are bosonic in nature, being harmonic oscillator-like wavefunctions with finite phase fluctuations enabled by the charging term, which acts as a kinetic energy in the phase representation. This kind of states correspond to Andreev replicas such as the higher doublet in Fig. \ref{fig:2fluxanium}a.

When $E_L$ is lowered, there is a competition between the inductive energy, a parabolic potential in phase representation, and the potential produced by the charge transfer terms. In the tunneling regime this potential is simply $-E_J\cos\hat{\phi}_{12}$, and the wavefunction may extend over two minima when $\phi_e=\pi$ and $E_J\sim E_L$, or several of the minima when $E_L\ll E_J$. In the latter case, if the kinetic energy $E_C$ is large compared to $E_J$, the wavefunction exceeds the maxima of the cosine and the phase delocalizes broadly; otherwise, the wavefunction localizes at several of the phase potential minima, and represents a superposition of discrete current states encircling the loop. In the junction with internal structure, the Andreev level introduces an additional degree of freedom. The limit $|\xi|\gg\Gamma$ recovers the previous tunneling regime where the states involve mainly just the lower energy Andreev state and the higher one is just virtually occupied to allow the flow of charge. However, in the resonant limit, both Andreev states participate. They are always degenerate at $\phi_e{=}\pi$ and if $E_L$ is lowered, their phase dependence decreases, providing a disjoint basis for a protected qubit insensitive to fluctuations in $\phi_e$ \cite{Caceres2022,Vakhtel2023flux}. The evolution with decreasing $E_L$ is shown in Fig. \ref{fig:2fluxanium}a,b,c where the black lines refer to a situation a bit out of resonance and the gray lines to a situation of full resonance, which maintains the Andreev degeneracy for any value of $E_L$.

In Figs. \ref{fig:2fluxanium}d,e we show the matrix elements $|\bra{0}\partial H / \partial \lambda \ket{1}|^2$ associated to noise in $\lambda{=}\phi_e$ ($\lambda{=}\xi$) between the two lowest energy states with solid (dashed) lines for the parameters in the panels a to c. There is a reduction in the matrix elements magnitude with the decrease of $E_L$ as the states spread over $\phi_{12}$ and become less sensitive to the Andreev degrees of freedom. In general, the protection against $\phi_e$ noise is enhanced when approaching resonance as the states share their structure in $\phi_{12}$ but are orthogonal in the Andreev sectors. In contrast, the noise on the parameters $\xi,\delta\Gamma$ that move away from resonance is more harmful in that regime as it produces dephasing due to the linear splitting of the energy levels. Finally, there is the issue of the phase drop distribution. In Fig. \ref{fig:2fluxanium}d,e we have used a symmetric drop in the couplings with the central level ($\alpha = 0$), as expected in an homogeneous field situation with almost symmetric coupling strengths $\Gamma_{1,2}$. However, a finite $\alpha$ would boost the effect of the $\phi_e$ noise because the pair tunneling terms would start to contribute to $\partial H/\partial \phi_e$.

\begin{figure*}[t]
    \centering
    \includegraphics[width=1\linewidth]{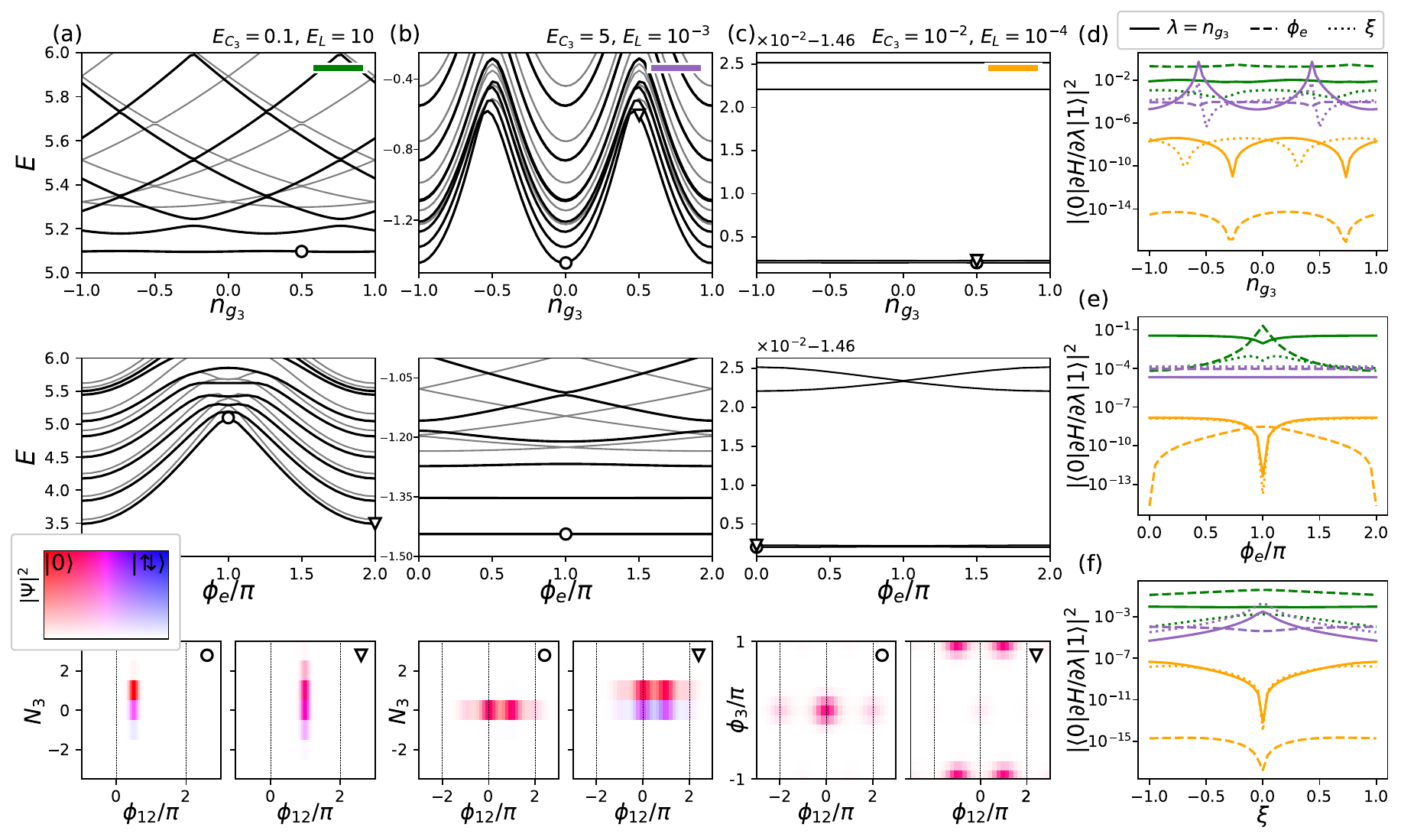}
    \caption{(a-c): Energy dispersion over the charge offset $n_{g_3}$ (top panels) and the external phase $\phi_e$ (middle panels) in the loop configuration where a third terminal is a charge island, and wavefunctions indicated with markers (bottom panels -- the coloured inset is a legend that represents the participation of the central level in the wavefunction). Gray lines indicate the resonant case and black ones a situation a bit out of resonance. In panel (a), the island is in the transmon regime (delocalized $\hat{N}_3$) and the loop is close to a phase biased regime (well localized phase, parameters of Fig. \ref{fig:2fluxanium}a and $\Gamma_3{=}1$). In panel (b), the island is in the Cooper pair box regime (localized $\hat{N}_3$, with only two states participating in the global wavefunctions), and the loop in a fluxonium regime [delocalized phase, parameters of Fig. \ref{fig:2fluxanium}b and $\Gamma_3{=}1$]. This is similar to a \textit{bifluxon}. In panel (c), the island has delocalized charge and the loop has delocalized phase (the resonant condition has a similar spectrum and it is not shown). This is similar to a \textit{0-}$\pi$ qubit -- the wavefunction is displayed in the $\ket{\phi_{12},\phi_{3}}$ basis to show that the wavefunctions of the two lowest states are disjoint. Parameters: $\Gamma_1{=}0.4$, $\Gamma_2{=}0.6$, $\Gamma_3{=}1$ and $\xi{=}0.6$ (energies in units of $E_{C_{12}}$). (d-f): Matrix elements associated to noise in $n_{g_3}$, $\phi_e$ and $\xi$ for the non-resonant parameters in (a-c) in the circle markers, and indicating each case with the colours in the (a-c) corners. \label{fig:3fluxjamon}}
\end{figure*}

\subsection{Three terminals: one loop and one island}

In the previous subsection we have seen that a low inductive potential favours the insensitivity to dephasing produced by $\phi_e$ noise. However, in order to achieve the protection against decoherence it is necessary to have two states that are disjoint over a different degree of freedom, such that $\partial H/\partial \phi_e$ does not connect them. These two states were provided by the fermionic structure of the junction in the resonant limit, which fulfils this purpose but introduces new sources of noise \cite{Caceres2022,Vakhtel2023flux}. We explore now the situation where a third terminal with charging energy is introduced (Eq. (\ref{eq:HT_loop_and_island})), and how it can provide disjoint states of a different origin. 

We begin in Fig. \ref{fig:3fluxjamon}a by connecting the loop with two terminals in the phase localized regime of Fig. \ref{fig:2fluxanium}a to an island in the transmon regime (delocalized charge). In the phase biased limit we have $E_L{\rightarrow} \infty$, $\hat{\phi}_{12}{\rightarrow} \phi_e/2$ and the configuration is similar to the one in Sec. \ref{sec:island.2}, where the central region is connected to one charge island and to one terminal without charging energy. In this analogy, $\Gamma_3$ is the coupling that shifts the occupation number of the island, and the term $\Gamma_{12}(\phi_e)=\Gamma_1 e^{-i\phi_e/2} + \Gamma_2 e^{i\phi_e/2}$ is the coupling that connects the central level states at a fixed island occupation, which is now tunable by $\phi_e$. This arrangement connects all sites in the resulting $\ket{N_3,n_d}$ chain and, as a result, the occupation of the island is determined by the combination of all the couplings, not only $\Gamma_3$. For instance, the wavefunction marked with a circle in Fig. \ref{fig:3fluxjamon}a has a lower charge dispersion that the one with a triangle because at $\phi_e=\pi$ the effective coupling $\Gamma_{12}(\phi_e)$ is reduced.

In Fig. \ref{fig:3fluxjamon}b we combine the phase extended regime of Fig. \ref{fig:2fluxanium}c in the loop with the island in the Cooper pair box limit (strongly localized charge, thus, it can be described with the two number states of lower energy). This situation benefits of the $\phi_e$ insensitivity already discussed in the previous subsection, but now the central region and the island hybridize into essentially $4$ levels. The phase potential they generate can produce disjoint states in the limit where only one level in the island (central region) contributes, and thus the central region (island) may be set at resonance. The first case is similar to the configuration of the previous subsection, and the second one describes the tunneling regime in the central region while two island states play the role of an effective resonant dot (discussed in next section and in Appendix \ref{ap:island2levels}). The protection of this configuration is analogous to that of a bifluxon qubit \cite{Kalashnikov2020}. 

In Fig. \ref{fig:3fluxjamon}c, we set the protected regime in each degree of freedom. The wavefunctions are delocalized over $\hat{\phi}_{12}$ and $\hat{N}_3$ (fluxonium and transmon-like), producing flat transition energies over the parameters $n_{g_3}$ and $\phi_e$, so the qubit states are protected against dephasing by the associated charge and flux noise. The protection against relaxation occurs because the states have disjoint wavefunctions (see lower panels for the wavefunctions and panels (d) to (f) for the matrix elements associated to noise on $n_{g_3}$, $\phi_e$ and $\xi$ for the settings in panels a to c). This disjointness is only observed in the space $\ket{\phi_{12},\phi_3}$, but the fact that the wavefunctions overlap in the space $\ket{\phi_{12},N_3}$ does not mean that it is not protected against noise on $n_{g_3}$, as the associated operator $\hat{N}_3 = -i\partial_{\phi_3}$ is local in the phase coordinate. The same applies to a possible charge noise in the loop with $\hat{N}_{12} \rightarrow \hat{N}_{12} {-} n_{g_{12}}$ and to noise in other parameters with local associated operators. Thus, the protection of this configuration is analogous to that of a $0-\pi$ qubit \cite{Gyenis2021_exp0pi}. This and the previous connection with protected qubits designed in multimode circuits made out from several traditional elements will be more easily discussed in the next section when focusing on the tunneling regime.

\section{Cotunneling limit}
\label{sec:cotunneling}

At certain limits, a non-tunnel Josephson junction embedded in a circuit can be described as a lumped element, i.e., an effective tunnel junction [Fig. \ref{fig:sketch}c, Sec. \ref{sec:island.2} and Appendix \ref{ap:adiabatic}], which is characterized by the phase dispersion of the lowest energy Andreev state. This adiabatic approximation in a multiterminal situation provides an interpretation in terms of processes that distribute pairs between the leads, i.e. a combination of lumped junctions between each terminal. For example, in a three-terminal configuration with time reversal symmetry (e.g. in Fig. \ref{fig:PhaseBiasedTrijunction}), the ground state can be written as 
\begin{equation*}
E_{0}(\phi_1,\phi_2) = \sum_{n_1,n_2 \in \mathbb{Z}} A_{n_1,n_2} \cos \left(n_1\phi_1{+}n_2\phi_2\right),    
\end{equation*}
where the form ${\propto} \cos \left(n_1\phi_1{+}n_2\phi_2\right)$ can be associated to processes where several Cooper pairs from one lead split into the other two, and $A_{n_1,n_2}=A_{-n_1,-n_2}$. This association may be done by defining the phase variables in correspondence with their operator counterparts in Sec. \ref{sec:island}. There, by using the reduced basis $\ket{N_1,N_2}$, a tunneling from terminals $1 \rightarrow 3$ ($2 \rightarrow 3$) is produced by the traslation operator $e^{-i\hat{\phi}_1}$ ($e^{-i\hat{\phi}_2}$), and from $1 \rightarrow 2$ by $e^{-i\hat{\phi}_1+i\hat{\phi}_2}$. In this way, the phase variables recover their {\it operator} nature conjugate to charge-like variables in the resulting effective Hamiltonian $H^{\textrm{eff}}_J(\hat{\phi}_1,\hat{\phi}_2)$ .

In particular, if we consider only single pair tunneling processes, the effective tunneling Hamiltonian is 
\begin{equation*}
H^{\textrm{eff}}_J = - E_{J_{13}}\cos \hat{\phi}_1 - E_{J_{23}}\cos \hat{\phi}_2 - E_{J_{12}}\cos{(\hat{\phi}_1{-}\hat{\phi}_2)},
\end{equation*} 
with $E_{J_{13}}{=}{-}2A_{1,0}$, $E_{J_{23}}{=}{-}2A_{0,1}$ and $E_{J_{12}}{=}{-}2A_{1,-1}$. We use this case to analyze the following limits.

\subsection*{Trijunction in an island configuration}
In the harmonic limit of an island configuration ($E_{C_\nu}$'s $\ll E_{J_{\mu\nu}}$'s), the wavefunctions localize in phase so we can approximate
$H_J^{\textrm{eff}} \approx (E_{J_{13}}+E_{J_{12}}) \hat{\phi}_1^2/2 + (E_{J_{23}}+E_{J_{12}}) \hat{\phi}_2^2/2 - E_{J_{12}}\hat{\phi}_1\hat{\phi}_2$. The charge offsets $n_{g_\nu}$'s have no effect in this limit (the charge becomes continuous and the $n_g$ a gauge freedom \cite{Devoret2021}) and we can write $H^{\textrm{eff}}_{\textrm{harm}} = \vec{N}^T \mathbb{C} \vec{N} + \vec{\phi}^T \mathbb{J} \vec{\phi}$ where $\vec{N}=(\hat{N}_1,\hat{N}_2)^T$, $\vec{\phi}=(\hat{\phi}_1,\hat{\phi}_2)^T$ and

\begin{align}
    & \mathbb{C} {=} \begin{pmatrix}
       E_{C_1} {+} E_{C_3}  & \hspace{-0.35cm} E_{C_3} \\
       E_{C_3}  & \hspace{-0.35cm} E_{C_2} {+} E_{C_3}
    \end{pmatrix}\hspace{-0.1cm}, \hspace{-0.1cm}
    & \mathbb{J} {=} \frac{1}{2}\hspace{-0.1cm}\begin{pmatrix}
       E_{J_{13}} {+} E_{J_{12}}  & \hspace{-0.35cm} {-}E_{J_{12}} \\
       {-}E_{J_{12}}  & \hspace{-0.35cm} E_{J_{23}} {+} E_{J_{12}}
    \end{pmatrix},
    %\nonumber
    \label{eq:matricesCJ}
\end{align}

\noindent
such that the Hamiltonian can be diagonalized into two harmonic modes  $H^{\textrm{eff}}_{\textrm{harm}} = \sum_{\nu} \omega_\nu (a^\dagger_\nu a^{}_\nu + 1/2)$ (see details and extensions provided in Appendix \ref{ap:sim_diag}).

In the limit where one island is in a Cooper pair box regime (e.g. $E_{C_3}\gg E_{J_{13}},E_{J_{23}}$), that island can be described with a two level model, one for each of the lowest energy number $N_3$ occupations. As elaborated in Appendix \ref{ap:island2levels}, it maps to the model of the Josephson-Andreev junction with two terminals in Sec. \ref{sec:island} when $E_{J_{12}}=0$ (the terminal $3$ would play the role of the central level). %When the other two terminals are in the transmon regime, if the coupling $E_{J_{12}} \ll E_{J_{13}},E_{J_{23}}$ its effect can be treated as a perturbation on the harmonic-like states defined by the bare effective JAnction; and if $E_{J_{12}} \gg E_{J_{13}},E_{J_{23}}$, the JAnction coupling terms can be treated as a perturbation on the harmonic-like states defined by the $E_{J_{12}}$ coupling.

\subsection*{Trijunction with two terminals in a loop \\ and one island configuration}

If two terminals close themselves forming a loop, the model is similar to the one in Sec. \ref{sec:loop}, where the variables are the tunneled pairs between the leads of the loop $\hat{N}_{12}$ and the number of pairs $\hat{N}_3$ in the island. The effective Hamiltonian is $H_{12} + H_3 + H_c$, where
\begin{align}
    & H_{12} = E_{C_{12}} \hat{N}_{12}^2 + E_{J_{12}} \cos 2\hat{\phi}_{12} +  E_L(2\hat{\phi}_{12}-\phi_e)^2 \nonumber \\
    & H_3 = E_{C_3} (\hat{N}_3 - n_{g_3})^2 \nonumber \\
    & H_c = E_{J_{13}} \cos{(\hat{\phi}_{12} + \hat{\phi}_3)} + E_{J_{23}} \cos{ (\hat{\phi}_{12} - \hat{\phi}_3)} \, ,
\end{align}

\noindent
and the operator $e^{i2\hat{\phi}_{12}}$ transfers a pair from $1 \rightarrow 2$, $e^{i\hat{\phi}_{12}+i\hat{\phi}_3}$ from $1 \rightarrow 3$ and $e^{i\hat{\phi}_{12}-i\hat{\phi}_3}$ from $2 \rightarrow 3$. The Hamiltonian can be solved in the basis $\ket{\phi_{12},N_3}$, and contains two modes which can be harnessed to devise protected qubits similarly to multimode circuits designed with superconductor-insulator-superconductor (\textit{SIS}) tunnel junctions \cite{gyenis2021}. Two conventional limits are the following.

On the one hand, if $E_{J_{12}} \gg E_{J_{13}},\, E_{J_{23}}$, the Hamiltonian is similar to a ``$\cos 2\varphi$'' qubit as the potential in $\hat{\phi}_{12}$ is almost $\pi$ periodic \cite{Smith2020,Larsen2020}. Physically, the situation with only $E_{J_{12}}$ tunneling conserves the parity of tunneled pairs, generating two disconnected sets of states.

On the other hand, if $E_{J_{12}} = 0$ and $E_{J_{13}}=E_{J_{23}} := E_J'$, the coupling becomes
\begin{align*}
    H_c= E_J' \cos{\hat{\phi}_{12}}\cos{\hat{\phi}_3},
\end{align*}
recovering the typical expression for multimode circuits, such as the bifluxon or the $0{-}\pi$ qubit, at the corresponding parameter regime \cite{gyenis2021}.

\section{General Discussion and Overview}
\label{sec:overview}

In this section we summarize and compare the main features of the different circuits discussed in previous sections and provide some estimates for the expected qubit frequencies and decoherence rates. For this purpose we compile in Table \ref{tab:1} the relevant information regarding degrees of freedom, control parameters, and circuit noise sources for each configuration. We also highlight their correspondence to conventional superconducting qubits based on tunnel junctions or other interesting properties.

\renewcommand*{\arraystretch}{2.5}
\newcommand{\tb}[1]{\textbf{#1}}
\begin{table*}[t!]
\begin{ruledtabular}
%\begin{tabular}{l{1cm} | r{1cm} r{1cm} r{1cm} r{1cm}}
%\begin{tabular}{p{3cm} | p{1cm} p{2cm} p{1cm} p{1cm}}
\begin{tabular}{c c c c c}
\tb{Configuration} & \tb{Quantum variables} & \tb{Control parameters} & \tb{Circuit noise} & \tb{Highlights} \\
\hline
%\vspace{-0.4cm}
\begin{tabular}{@{}c@{}}
 \vspace{-0.7cm}\\
\includegraphics[width=0.12\linewidth]{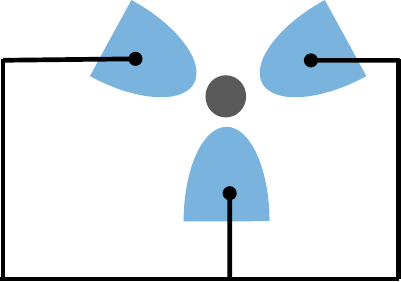} 
\vspace{-0.47cm}\\
 Perfect phase bias \end{tabular}&
$n_d \in \mathbb{Z}_2$ &
\begin{tabular}{@{}c@{}} $\phi_{1e}$, $\phi_{2e}$ \\ $\xi$, $\Gamma_\nu$ \end{tabular}
 &
Flux & 
\begin{tabular}{@{}c@{}}  \begin{tabular}{@{}c@{}} - Sensitivity of relaxation to \vspace{-0.45cm} \\
phase drop distribution \vspace{-0.37cm}\end{tabular}\\
- Appearance of Weyl points \vspace{-0.37cm} \\
- Connection with Bi-SQUID \end{tabular} \\ 
\hline

%\vspace{-1.4cm}
\begin{tabular}{@{}c@{}}
 \vspace{-0.7cm}\\
\includegraphics[width=0.14\linewidth]{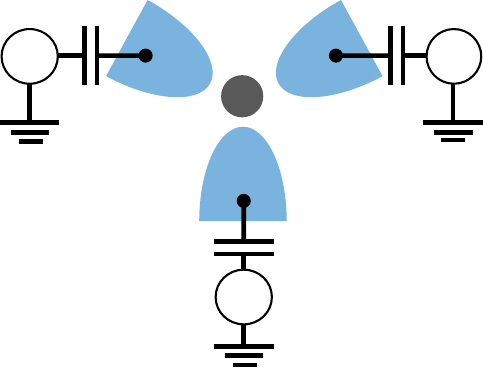} \vspace{-0.47cm}\\
 Charge islands \end{tabular}&
\begin{tabular}{@{}c@{}} $N_\nu \in \mathbb{Z}$ \; ($\phi_\nu \in [-\pi,\pi)$) \\ $n_d \in \mathbb{Z}_2$\end{tabular} &
\begin{tabular}{@{}c@{}} $n_{g_\nu}$ \\ $\xi$, $\Gamma_\nu$ \end{tabular} &
Charge &
\begin{tabular}{@{}c@{}} 
\begin{tabular}{@{}c@{}} - Mapping to a discrete lattice: \vspace{-0.43cm}\\
 platform for topology \end{tabular}    
\\ - Controllable disjointness\end{tabular} \\
\hline

\begin{tabular}{@{}c@{}}
 \vspace{-0.77cm}\\
\includegraphics[width=0.14\linewidth]{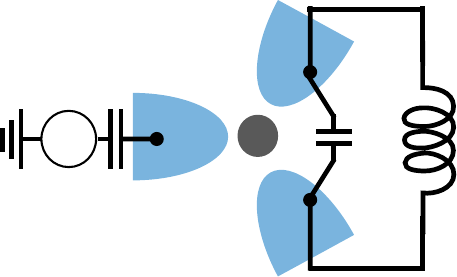} \vspace{-0.37cm}\\
 Hybrid charge/flux \end{tabular}&
\begin{tabular}{@{}c@{}} \vspace{-0.3cm} $N_{12} \in \mathbb{R}$, $N_3 \in \mathbb{Z}$ \\ ($\phi_{12} \in \mathbb{R}$, $\phi_3 \in [-\pi,\pi) $) \\ $n_d \in \mathbb{Z}_2$ \end{tabular} &
\begin{tabular}{@{}c@{}} $\phi_{e}$, $n_{g_3}$ \\ $\xi$, $\Gamma_\nu$ \end{tabular} &
Charge, flux &
\begin{tabular}{@{}c@{}} 
\begin{tabular}{@{}c@{}} - $0$-$\pi$ and bifluxon \vspace{-0.5cm}\\ in the tunnel limit \vspace{-0.5cm}\end{tabular} \\ 
\vspace{-0.5cm}$\Updownarrow$\\
\begin{tabular}{@{}c@{}} - Simultaneous noise protection to \vspace{-0.5cm} \\ relaxation and $n_g$, $\phi_e$-dephasing \end{tabular}
\end{tabular} \\
\end{tabular}
\end{ruledtabular}
\caption{Summary of the circuit configurations for three terminal junctions analyzed in this work, indicating the relevant quantum variables and control parameters within our modeling, the main circuit noise sources and the most significant properties. First row for the perfect bias case (Sec. \ref{sec:phase_bias}), second row for charge islands configuration (Sec. \ref{sec:island}) and third row for the mixed configuration including a loop and a charge island (Sec. \ref{sec:loop}). \label{tab:1}}
\end{table*}

The first row in this table corresponds to the triterminal junction with perfect phase-biasing considered in Sec. \ref{sec:phase_bias}. In this case the control parameters are the external fluxes $\Phi_{1e}$, $\Phi_{2e}$ determining the phase differences between the junction terminals. In this configuration the phases have no dynamics and we deal with a purely fermionic Andreev qubit. At the central level degeneracy point, $\xi{=}0$, this configuration is interesting due to the appearance of protected crossings (Weyl points) (see Fig. \ref{fig:PhaseBiasedTrijunction}). As the central level becomes more detuned, the energy bands become flat at the expense of increasing its transition energy. A similar spectrum has been shown to arise in a bi-SQUID circuit based on metallic tunnel junctions \cite{peyruchat2024}.

The second row in Table \ref{tab:1} corresponds to the charge-islands configurations discussed in Sec. \ref{sec:island}. The two terminal device in the tunnel regime describes the gatemon for gate tunable weak links and when the third lead is connected, the total charge delocalizes in a plane of integer quantum variables. We find an interesting regime [Figs. \ref{fig:3T}c,d] where the wave-functions correspond to the charge chains of two islands coupled through the third terminal, which thus controls their disjointness.  Interestingly, the mathematical (``crystalline'') structure of the charge-island configurations can be mapped into topological band models.

Finally, the third row in Table \ref{tab:1} corresponds to the case where two terminals are shunted by a loop with significant inductive energy. In the two-terminal case we explore the transition from an almost localized phase degree of freedom to a fluxonium-like regime. 
The most interesting case corresponds to this later one, where there is simultaneous protection to decoherence in flux-noise due to the phase delocalization [Fig. \ref{fig:2fluxanium}c] and a disjoint support with respect to the Andreev character \cite{Caceres2022,Vakhtel2023flux}. When the third lead is connected, we explored a hybrid configuration in which a charge island is coupled through the central level to a flux-loop with nonzero inductance. We compare a transmon-(flux qubit)-like regime [Fig. \ref{fig:3fluxjamon}a], a (Cooper pair box)-fluxonium-like regime [Fig. \ref{fig:3fluxjamon}b], and a transmon-fluxonium-like regime [Fig. \ref{fig:3fluxjamon}c] with simultaneous delocalization in the charge and phase variables (flat bands in the parameter space) and disjoint wavefunction support. This is an interesting proposal because it can realize a 0-$\pi$ qubit, whose explicit mapping to the conventional implementation we show in the tunnel limit in Sec. \ref{sec:cotunneling}.

\subsection*{Poisoning}
\begin{figure}[t]
    \centering
    \includegraphics[width=1\linewidth]{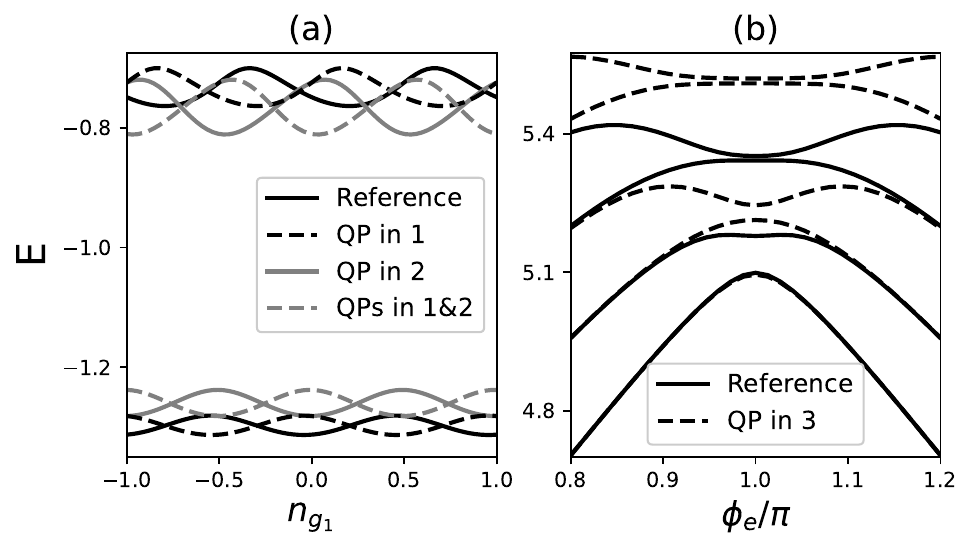}
    \caption{States with quasiparticle poisoning in the islands. (a) Three terminal configuration of Fig. \ref{fig:3T}b with two charge islands. (b) Charge-flux hybrid configuration of Fig. \ref{fig:3fluxjamon}a at $n_{g_3} = 1/4$. \label{fig:QPpois}}
\end{figure}

We discuss here the effects of quasiparticle (QP) poisoning, that takes place when unbound electrons with energies over the gap tunnel from one region to another \cite{Glazman2021}. Though these events may occur at long timescales compared with some of the present qubit decoherence times, they can become limiting factors when other decoherence sources are optimized, and many strategies are being developed to decrease the density of excess QPs as it follows a non thermal distribution highly dependent on the implementation. We distinguish two kinds of poisoning contributions according to the region where the quasiparticle infiltrates into. 
As a first scenario, poisoning on the charge islands or between capacitively coupled leads produces a shift $N_\nu \rightarrow N_\nu \pm 1/2$, which is common in two-terminal devices but more complex in multiterminal ones. As we show in Fig. \ref{fig:QPpois}, poisoning leads to the appearance of additional states for the case of three terminal devices. In the configuration with two charge islands [panel (a)], poisoning may occur in any of them and this provides four families of transitions \cite{Wills2022}. In the situation where all terminals are islands with finite charging energy, poisoning events from the environment into the system do modify the spectrum even if they appear in pairs, because states with different total charge are not equivalent (\textit{e.g.} Fig. \ref{fig:3T}a). In the configuration with a loop and an island (panel b), the states produced by poisoning in the island depend both on $n_{g_3}$ and $\phi_e$. The poisoning between the leads in the loop does not affect the spectrum as the shunting renders the charge difference continuous and an static offset charge becomes a gauge freedom \cite{Koch2009}, but it does generate relaxation \cite{Glazman2021,Pop2014}. 

As a second possibility, poisoning on the junction central level, in contrast with the previous case, produces a major rearrangement of the wavefunction as the quasiparticle mediates the supercurrent. This cannot be described with a trivial shift of a parameter, but requires a model that includes the effect of the excess quasiparticles. Within our infinite gap approximation, isolated QPs at the central region are not taken into account and thus this odd state is not included. Including such effects could be possible by means of models beyond the infinite gap approximation like the ones used in Refs. \cite{Pavesic2024,Baran2024}, which can be undertaken in future studies.

\begin{figure*}[t]
    \centering
    \includegraphics[width=2\columnwidth]{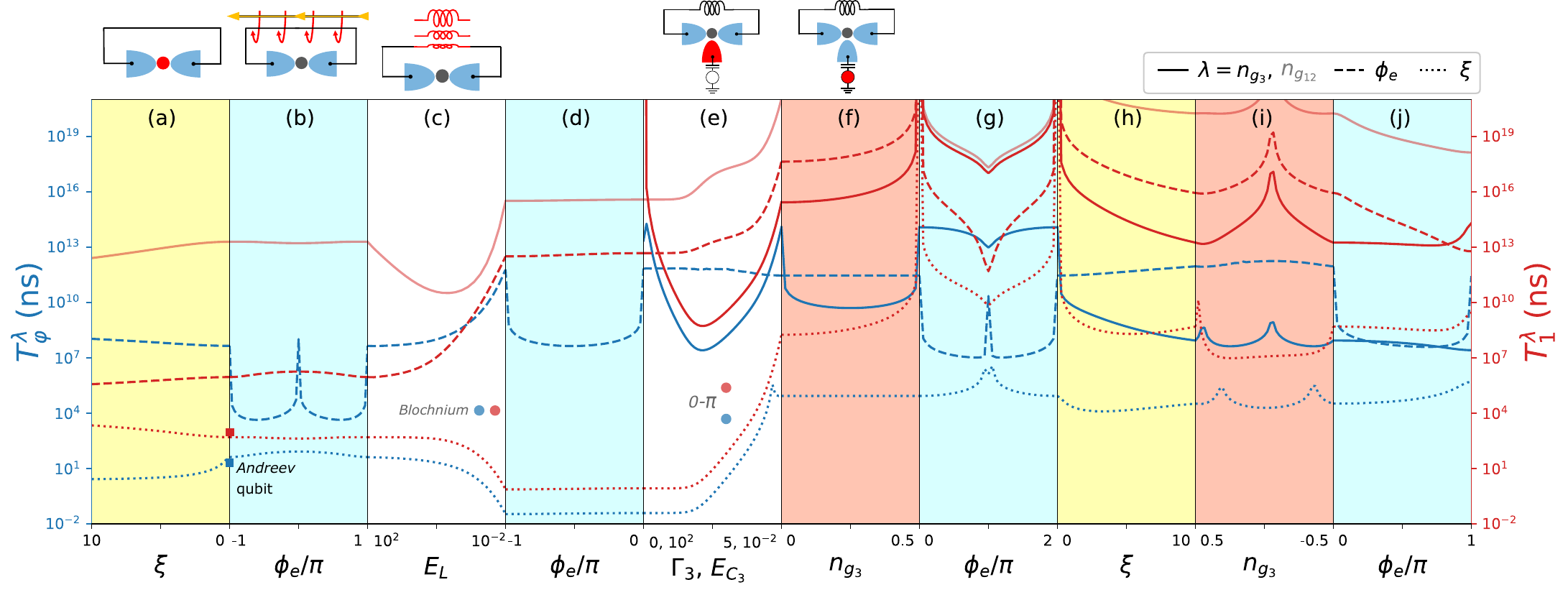}
    \caption{Dephasing (blue lines) and decoherence (red lines) estimated times for different configurations with a loop, while sweeping the position of the central level (light yellow background), the external flux (light cyan), the island offset charge (light orange) and other parameters to cross over between different regimes (white). (a,b) Perfect phase bias with two terminals. (c) Transition to introduce phase fluctuations until the fluxonium-like regime is reached in (d). (e) Coupling with the charge island and transition to $0$-$\pi$-like protected regime in (f,g). (h) Transition deep into tunnel regime and evolution with island gate and external flux in (i,j). Parameters are given in the text, units in GHz. Decoherence times of the Andreev level qubit used for the choice of some parameters \cite{Metzgerthesis} are indicated with a squares; other markers indicate a Blochnium and a $0$-$\pi$ implementation in conventional circuits just for illustrative purposes \cite{Pechenezhskiy2020,Gyenis2021_exp0pi}.
    \label{fig:decoherence}}
\end{figure*}

\subsection*{Relaxation and dephasing}
We now discuss possible sources of decoherence and relaxation. On the one hand, the measurement device used to probe the quantum states of the circuit, which can be a readout microwave resonator, will induce some relaxation (by means of Purcell effect, or dressed dephasing \cite{Tosi2024}) and dephasing (through measurement-induced dephasing). This contribution depends strongly on the design and can be mitigated by different well developed strategies (like Purcell filters). 
On the other hand, the noise on the system parameters produces an inherent decoherence, that is to be minimized by the choice of the architecture of the device and its operation point. In circuits with conventional tunnel junctions those parameters are the offset charge, the external flux and the Josephson coupling. But in junctions with an Andreev structure, each fermionic degree of freedom (in the single level model controlled by parameters $\xi$ and the $\Gamma_\nu$'s) may also experience its own noise. Other source of decoherence is the relaxation by dielectric loss, which depends on the design and the materials \cite{Martinis2005,Pop2014,Sun2023,Zhang2021}. 

There are mainly two kinds of noise affecting a parameter: 1/f and quantum Johnson-Nyquist noise \cite{Krantz2019}. For instance, a background of two level systems produces 1/f fluctuations in the offset charges, while the coupling with a gate is responsible for its Johnson-Nyquist noise $\propto \omega\coth{\omega/2k_BT}$. Precise models for the spectral noise from these sources depend on the particularities of the system. Thus, here we use a simple expression able to reproduce typical behaviour of the spectral noise. As the crossover between noises occurs typically in the range of GHz both for offset charge and flux \cite{Krantz2019,Astafiev2004,Yan2016,Quintana2017} this form can be chosen as
\begin{equation*}
    S_\lambda(\omega) = 2\pi A_\lambda^2 \left( \frac{1}{\omega} + \frac{\eta_\lambda \omega}{1\textrm{GHz}^2} \coth{ \frac{\omega}{1\textrm{GHz}} } \right),
\end{equation*}

\noindent
where $A_\lambda$ is a noise amplitude, $\eta_\lambda$ is a factor adjusting the Johnson-Nyquist noise and we have used a temperature $T \sim 25 \rm{mK} \approx 0.5 \rm{GHz}$ and a symmetrized spectral noise such that the relaxation (depolarization) rate $1/T_1^\lambda = S_\lambda(E_{01}) |\bra{0}\partial H / \partial \lambda \ket{1}|^2$ includes both emission and absorption at the qubit frequency (we disregard excitation processes to higher energy states, but in the disjoint regime these may be relevant \cite{Groszkowski2018}). We use typical noise amplitudes for charge and flux noise $A_{n_g} {\sim} 10^{-4}$, $A_{\phi_e} {\sim} 10^{-5} $, and $\eta_{n_g} {\sim} 0.01$, $\eta_{\phi_e} {\sim} 10$ \cite{Ithier2005,Koch2007,Krantz2019,Janvier2015}. For the noise in the Andreev structure, we note that few experiments exist on mesoscopic Josephson junctions providing an analysis for relaxation and dephasing rates, so the accuracy of this discussion is limited. However, we try to give some reasonable estimation, for example, from the data in Ref. \cite{Metzgerthesis}, where there is a single central gate, which mainly controls $\xi$ (also $\Gamma$ but with lower lever arm). Since there are fewer experiments which implement several gates, and none of them characterizes the decoherence, we consider just the effect on $\xi$. We use the same expression as for the $n_g$ noise corresponding to a gate electrode, with the prefactor $A_\xi {\sim} 0.04$ GHz, which provides the order of magnitude of the dephasing and relaxation times found in nanowire junctions ($T_1 {\sim} 1\mu$s, $T_\varphi {\sim} 50$ns in Ref. \cite{Metzgerthesis}, $T_1 {\sim} 10\mu$s in Ref. \cite{Hays2018}), also qualitatively similar in a point contact \cite{Janvier2015}).

For the dephasing, we account only for the noise close to $\omega=0$, \textit{i.e.}, the 1/f contribution, which produces a typical time $T_\varphi^\lambda$

\begin{equation*}
    \left(\frac{1}{T_\varphi^\lambda}\right)^2 =  c_1 A_\lambda^2 \left(\frac{\partial E_{01}}{\partial \lambda}\right)^2 + c_2 A_\lambda^4 \left(\frac{\partial^2 E_{01}}{\partial \lambda^2}\right)^2, 
\end{equation*}

\noindent
where $c_{1,2}$ are coefficients that depend on the cutoff frequencies. We use typical values $c_{1,2} \sim 30,1200$ \cite{Groszkowski2018,Ithier2005} (but we do not include those cutoffs for the relaxation calculation).

We first discuss briefly the charge islands configurations (Sec. \ref{sec:island}). The most amenable two terminal configuration for hosting a qubit is the \textit{transmon} regime, when the ratio between effective Josephson coupling $E_J^{\textrm{eff}}$ and the charging energy is large enough to reduce the transition dispersion with the charge offset but not as much as to lose significant anharmonicity. This suppression of charge dispersion is enhanced when the Andreev structure approaches the resonant condition \cite{Bargerbos2020,Kringhoj2020,Vakhtel2023}. The relaxation from charge noise is negligible \cite{Ithier2005,Koch2007}. The Andreev structure provides $E_J^{\textrm{eff}}$ (Appendix \ref{ap:adiabatic}), which we analyze for simplicity at $\delta\Gamma=0$ such that $E_J^{\textrm{eff}} = \Gamma^2/\sqrt{\Gamma^2+\xi^2}$. In conventional transmons of oxide tunnel barriers, the fluctuations of the Josephson coupling are produced by spatial reconfigurations of ions inside the junction, among other mechanisms \cite{Koch2007}. In our model of the Josephson-Andreev junction, we expect additional noise from the gates that control the position of the level and the couplings. At fixed $\Gamma$, dephasing by $\xi$ behaves similarly to the dephasing in the perfect flux bias since $E_{01} \sim 2\sqrt{E_{C_1} E_J^{\textrm{eff}}}$. It has a minimum at $\xi=0$, then increases till $\xi\approx \Gamma/\sqrt{2}$ (using the harmonic limit) and then decreases with a tail $\sim \xi^{-4}$. Another comparison can be made by fixing $E_J^{\textrm{eff}}$ by moving $\Gamma$ accordingly. In this case, the $\xi$-dephasing maximum is at $\xi \approx 4\sqrt{3}E_J^{\textrm{eff}}$. Relaxation is negligible as can be noted in the harmonic and adiabatic limits $\bra{0}\cos \hat{\phi} \ket{1} = 0$ by means of parity arguments. Regarding tunability, experiments on gatemons tuned close to resonance have been performed in the regime $E_{C_1}{<}E_J^{\textrm{eff}}{\sim}\Gamma/8{\sim} 5$GHz, with qubit frequencies in the order of a few GHz \cite{Bargerbos2020,Kringhoj2020}. In the case of three terminals in the charge island configuration, as discussed in Sec. \ref{sec:islands.discussion}, having a low $\Gamma_3$ is analogous to the Cooper Pair Box-regime and, when all $\Gamma_\nu$ are similar, to a transmon-like regime but in a larger space. This configuration is interesting for fundamental reasons more than for qubit applications. 

We now discuss the loop configurations (Secs. \ref{sec:phase_bias} and \ref{sec:loop}), and show the estimated relaxation and dephasing times in the hybrid charge-flux configuration [Fig. \ref{fig:sketch}f] for different regimes [Fig. \ref{fig:decoherence}]. First, we set $\Gamma_3=0$ and $E_L=100$~GHz to recover the perfect phase bias configuration with terminals [panels a and b]. When decreasing $\xi$ towards the sweet spot [panel a], the associated dephasing rate, which dominates the decoherence, is reduced at the expense of an increase in the associated relaxation rate. The same occurs (b) with the $\phi_e$ noise towards its dephasing sweet spot. In this sweep, the qubit frequency oscillates between $10$~GHz and $20$~GHz reproducing the measurements in Ref. \cite{Metzgerthesis}, with $\Gamma =10$~GHz and $\delta\Gamma=5$~GHz. We keep these parameters whose realization is less challenging than the perfectly resonant limit. In (c), we increase (logarithmically) the inductance up to $E_L = 10$~MHz, in the order of the limit of what is feasible \cite{Pechenezhskiy2020}, allowing for phase fluctuations. As it is possible to control the capacitive couplings in the terminals, we use a large charging energy $E_{C_{12}} = 50$~GHz. Thus, the result is analogous to a Blochnium \cite{Pechenezhskiy2020} as $E_{C_{12}} > E_J^{\textrm{eff}} \sim 1$~GHz, with $\phi_e$-dephasing sweet spots at $\phi_e=0,\pi$ and large $\xi$-induced decoherence (d). The qubit frequency in this panel and the following ones is reduced, in the order of $10$~MHz. When we connect the third lead (e) changing $\Gamma_3$ from 0 up to $\Gamma_3{=}5$~GHz, while decreasing (logarithmically) the charging energy $E_{C_3}$ of the island down to $0.01$~GHz -which was set at a large value to avoid other states ($100$~GHz)-, we arrive at a protected $0$-$\pi$-like regime, which is tested over $n_{g_3}$ and $\phi_e$ (f,g). The relaxation from all parameters (even $\xi$) is reduced because the states become considerably disjoint, particularly in the point $n_{g_3}{=}1/2$, $\phi_e {=} 0$, though that extreme protection is lost at finite $\xi$ (h). When the central region is brought more into the tunnel regime (h,i,j), the $0$-$\pi$-like protection is maintained. We note that even if the Andreev structure confers an additional layer of complexity, the relevant parameters to achieve this kind of disjointness --that occurs in the circuit bosonic variables-- are the relations between the effective Josephson couplings and the charging energies. Also, the dominant effect of slow dephasing by $\xi$ noise can be mitigated by several techniques, such as Hahn echo and other dynamical decoupling sequences. In addition, we expect no decoherence from the coupling with other circuit {\mbox -spurious-} modes, which are present in some implementations of multimode protected circuits with conventional tunnel junctions but are not present in our charge-flux hybrid configuration  \cite{Groszkowski2018,Dempster2014,Smith2020,gyenis2021,Gyenis2021_exp0pi}. Finally, let us mention that the rather long calculated timescales are the result of the choice of the parameters in a  ``hard'' regime, similar to calculations for $0$-$\pi$ implementations in conventional circuits \cite{Groszkowski2018,gyenis2021}. 

\section{Conclusions}
\label{sec:conclusions}

Mesoscopic Josephson junctions host localized states with phase-dependent energy which can be probed in spectroscopy and manipulated coherently using microwave pulses. The number of these states, their spin texture and their energy are sensitive to internal as well as control parameters such as electric fields applied with a gate voltage, magnetic fields, or strain. 
Compared to standard tunnel junctions, the energy of the fermionic excitations may become of the same order of the collective bosonic modes of quantum circuits that include capacitors and inductors. This ``mesoscopic embbeding'' constitutes a promising experimental and theoretical research topic with opportunities to develop hybrid fermionic-bosonic devices more immune to decoherence than conventional ones. Additionally, on a fundamental level, these devices raise questions on the proper modeling and on the correct quantum circuit rules for their combination. In this work we have addressed these questions taking a three-terminal junction connected in different configurations as a model system.

While exploring the energy spectrum and the wavefunctions of these Josephson-Andreev junctions for different circuit configurations we have identified regimes of charge and/or flux noise protection against dephasing and/or relaxation. We have discussed different limiting cases for each configuration to recover known results, and for the multiterminal configurations we have shown their connection with other kinds of protected qubits such as the bifluxon or the $0{-}\pi$ qubits obtained using standard tunnel junctions arrays. This fact highlights how the multimode character of these circuits can be emulated thanks to the connectivity of the multiterminal ones, with the additional value of their higher degree of tunability.

Regarding the modeling, the nuance of the gauge \textit{choice} for time dependent situations, which includes the considerations about the noise, emerges as an important issue to care about. Though we provide some orientation for this choice in our problem, its strong effect suggests that experimental guidance would be required for more conclusive estimations. Additionally, several open issues remain to be investigated, such as the robustness with respect to quasiparticle poisoning, which requires a more elaborate model for the superconducting leads, or the design of suitable operation protocols. We expect that the present work could motivate further experimental and theoretical efforts along these lines of research.  

\begin{acknowledgments} 
We thank G. O. Steffensen, A. Zazunov, L. Arrachea, Ch. Strunk and N. Paradiso for useful discussions; and J. Berger for details on the possible experimental realization of multiterminal hybrid junctions. We are also grateful to J. J. C\'aceres, E. Flurin, M. F. Goffman, H. Pothier and C. Urbina from the Quantronics group for having shared their ideas on fermionic-bosonic qubits with us. A.L.Y. and F.J.M. acknowledge support from the Spanish AEI through grants PID2020-117671GB-I00, TED2021-130292B-C41 and through the “Mar\'ia de Maeztu” Programme for Units of Excellence in R\&D (Grant No. MDM-2014-0377) and from Spanish Ministry of Universities (FPU20/01871). Support by EU through grant no. 828948 (AndQC) is also acknowledged. Leandro Tosi acknowledges the Georg Forster Fellowship from the Humboldt Stiftung. 
\end{acknowledgments}

\appendix
\section{Effective Hamiltonian \\ in phase biased conditions}
\label{ap:gauge}
Our starting point is a single-level model coupled to superconducting leads, described by a Hamiltonian of the form
\begin{equation*}
H = \sum_{\sigma} \epsilon d^{\dagger}_{\sigma} d_{\sigma} + H_T + \sum_{\nu} H_{\nu} \;,
\end{equation*}
where $H_{\nu}$ correspond to the different terminals ($\nu$ ranging from 1 to 3 in the trijunction case) and
\[ H_T = \sum_{\nu,\sigma,k} t_{\nu} e^{-i\phi_{\nu}/2} d^{\dagger}_{\sigma} c_{\nu,k\sigma} + \mbox{h.c.} \] 
where $\phi_{\nu}$ is the superconducting phase on each lead. 

Using conventional field theoretical methods one can integrate out the leads, leading to an effective action on the dot
\begin{equation} 
S_{\text{eff}} = \int dt dt' \bar{\psi}_d(t) \left[ i\delta(t-t') \left(\partial_t - \epsilon \tau_z\right) - \hat{\Sigma}(t,t') \right] \psi_d(t') \;
\label{effective-action}
\end{equation}
where $\psi_d$ is the Grassman field associated with the dot Nambu spinor $(d_{\uparrow} \; d^{\dagger}_{\downarrow})^T$, and $\hat{\Sigma}(t,t')$ denotes the leads 
self-energy given by
\begin{equation*}
\hat{\Sigma}(t,t') = \sum_{\nu} t^2_{\nu} e^{-i\phi_{\nu}(t)\tau_z/2} \hat{g}(t-t') e^{i\phi_{\nu}(t')\tau_z/2} \;.
\end{equation*}

For the uncoupled leads Green functions $\hat{g}(t-t')$ we can use the BCS model in the wide band approximation, i.e.
\begin{equation*}
\hat{g}(t-t') = \int \frac{d\omega}{2\pi} e^{i\omega(t-t')} \left[\frac{-\omega \tau_0 + \Delta \tau_x}{W\sqrt{\Delta^2 - \omega^2}}\right] \;, 
\end{equation*}
where $W$ denotes the bandwidth. In the static case and in the absence of phase fluctuations (constant $\phi_{\nu}$) Eq. \ref{effective-action} can be written as
\begin{multline*} 
S_{\text{eff}} = \int d\omega \bar{\psi}_d(\omega) \Biggl[\omega - \epsilon \tau_z - \\
\left. \sum_{\nu} \Gamma_{\nu} \left(\frac{-\omega \tau_0 + \Delta \left(\tau_x \cos\phi_{\nu} - \tau_y \sin \phi_{\nu}\right)}{\sqrt{\Delta^2 - \omega^2}}\right) \right] \psi_d(\omega) \;,
\end{multline*}
where $\Gamma_{\nu} = t^2_{\nu}/W$. In the large gap limit, i.e. $\omega \ll \Delta$ the model is characterized by bound states at 
\begin{equation*}
E_A(\left\{\phi_{\nu}\right\}) = \sqrt{ \epsilon^2 + |\sum_{\nu} \Gamma_{\nu} e^{-i\phi_{\nu}}|^2} \,.
\end{equation*} 

On the other hand, in the presence of phase fluctuations Eq. \ref{effective-action} can be simplified following the lines of Ref. \cite{Zazunov2005}. For that purpose we approximate 
$\hat{g}(t-t')$ as
\begin{equation*}
\hat{g}(t-t') \simeq \delta(t-t') \frac{-i\partial_t \tau_0 + \Delta \tau_x}{W \Delta} \;, 
\end{equation*}
which leads to

\begin{multline*}
S_{\text{eff}} = \int dt  \bar{\psi}_d(t) \Biggl[ i\partial_t - \epsilon \tau_z - \\ 
 \left. \sum_{\nu} \Gamma_{\nu} \left(\tau_x \cos\phi_{\nu}(t) - \tau_y \sin \phi_{\nu}(t) - 
\tau_z \frac{\dot{\phi}_{\nu}(t)}{2\Delta} \right) \right] \psi_d(t) \;.
\end{multline*}

To remove the $\dot{\phi}_{\nu}$ terms we redefine $\psi_d(t) \rightarrow e^{-i\sum_{\nu} \tilde{\Gamma}_{\nu} \phi_{\nu} \sigma_z} \psi_d(t)$, where $\tilde{\Gamma}_{\nu} = \frac{\Gamma_{\nu}}{2\Delta}$, so that 
\begin{equation*}
S_{\text{eff}} = \int dt  \bar{\psi}_d(t) \left[ i\partial_t - \hat{H}_{\text{eff}} \right] \psi_d(t) \;,
\end{equation*}
where 
\begin{multline}
\hat{H}_{\text{eff}} = e^{-i\sum_{\nu} \tilde{\Gamma}_{\nu} \phi_{\nu} \tau_z} \Biggl[ \epsilon \tau_z + \\
\left. \sum_{\nu} \Gamma_{\nu} \left(\tau_x \cos\phi_{\nu} - \tau_y 
\sin \phi_{\nu} \right) \right] e^{i\sum_{\nu} \tilde{\Gamma}_{\nu} \phi_{\nu} \tau_z} \;.
\label{effective-hamiltonian}
\end{multline}

This last expression can be used as the effective Hamiltonian for the multiterminal quantum dot junction. In practice only two out of the three phases $\left\{\phi_{\nu}\right\}$ can be considered independent. Also, in the infinite gap limit one can disregard the phase prefactors in Eq. \ref{effective-hamiltonian}.

%\section{Projection of the higher Andreev band}
%\label{ap:downfolding}
%\input{appendices/downfolding.tex}
\section{Adiabatic approximation}
\label{ap:adiabatic}
When transforming from the number space to the phase space (for simplicity in a two-terminal configuration with one degree of freedom), it is convenient to interpret the latter as a discrete grid, that arises when the number space is truncated to an arbitrarily large interval $N{\in}[-N^M,N^M]$ with periodic boundary conditions, defining $\ket{\phi_n} = N_\#^{-1/2}\sum_{N} e^{iN\phi_n}\ket{N}$, with $N_\#=2N^M{+}1$ and $\phi_n{=}2\pi n/N_\#$. This is justified because at finite charging energy the wavefunction has no infinite extension in $N$. Then, the operator $\hat{N}$ can be thought as a translation in the $\ket{\phi_n}$'s, with $\hat{N}^2 \approx 2N_\#^2(1-\cos \hat{N}/N_\#)$. In the continuum limit this converts into the second derivative ${-}\partial_{\phi}^2$.

At each point $\phi$, the potential Hamiltonian $H_A(\phi)$ is equivalent to the one describing the phase biased configuration (Eq. \ref{Heffeven}), up to a \textit{gauge} transformation. The Andreev transformation $U(\phi)H_A(\phi)U^\dagger(\phi) = \mathbb{E}_A$ diagonalizes the states at each phase point, but then the \textit{kinetic} term $\sim E_{C}\hat{N}^2$ becomes non-diagonal in the neighbouring phase hoppings:
%\begin{align*}
%    (\hat{N}{-}n_g)^2 \rightarrow \mathbbm{1} {+} (\hat{N}{-}n_g)^2 {+} i \comm{U'(\phi) U(\phi)^\dagger}{ (\hat{N}{-}n_g) - i (\hat{N}{-}n_g) U(\phi) U'(\phi)^\dagger, 
%\end{align*}
\begin{align*}
    \hat{N}^2 \rightarrow \hat{N}^2 - i \acomm{U(\phi) U'^\dagger(\phi)}{\hat{N}} - (U(\phi)U'^{\dagger}(\phi))^2, 
\end{align*}

\noindent
where $U'(\phi) = \partial_\phi U(\phi)$, and it has been used that $\partial_\phi(U(\phi)U^\dagger(\phi))=0$. The adiabatic approximation consists in truncating the Andreev sector to its ground state (with energy equal to $\min{\{\mathbb{E}_A\}}$) and keeping the main kinetic term, which is the one with $\hat{N}^2$ since in the transmon-like regime the wavefunction is strongly delocalized in charge. We note that the Andreev transformation has distorted the number translation and that there is a gauge dependence similar to the one in Sec. \ref{sec:phase_bias}, that arises from the choice of the reduced number basis (the complete charging term would also depend on it). When restricted to the truncated Andreev sector, it has a similar role to a phase dependent charge offset and a phase potential, which becomes less noticeable as the charge delocalizes.

In this limit, the phase localizes and we can expand the effective potential into an anharmonic oscillator 
\begin{align*}
    H^{\textrm{eff}}_{\text{osc}} = E_C \hat{N}^2 + E^{\textrm{eff}_2}_{J} \hat{\phi}^2 + E^{\textrm{eff}_4}_J \hat{\phi}^4, 
\end{align*}
\noindent
with $ E_J^{\textrm{eff}_n} = \partial^n_\phi E_0(\phi) \bigr\rvert_{\phi{=}0}/n!$, which in the harmonic basis $\ket{m} {=} (a^{\dagger})^m\ket{0}/\sqrt{m!}$, with $a^{(\dagger)} = (E_J^{\textrm{eff}_2}/4E_C)^{1/4} \hat{\phi}_1 \mp i (E_C/4E_J^{\textrm{eff}_2})^{1/4}\hat{N}$ (see Appendix \ref{ap:sim_diag}), produces states with energies $E_m$ equal to
\begin{align*}
    2\sqrt{E_C E_J^{\textrm{eff}_2}}\left(m+1/2\right) + \frac{E_C E_J^{\textrm{eff}_4}}{4E_J^{\textrm{eff}_2}}\left(6m^2+6m+3\right), 
\end{align*}

\noindent
 up to first order perturbation theory. This yields a transition $E_{01}$ equal to the prefactors on $m$ (the one with the square root is referred with $\omega$ in the main text) and an anharmonicity $\alpha_{h} := E_{12}{-}E_{01}$ equal to twice the prefactors on $m^2$. Within the model of Sec. \ref{sec:phase_bias}, $E_J^{\textrm{eff}_2} = (\Gamma^2{-}\delta\Gamma^2)/8\sqrt{\Gamma^2{+}\xi^2}$ and $\alpha_h = E_C\left({-}1/4 + 3(\Gamma^2{-}\delta\Gamma^2)/16(\Gamma^2{+}\xi^2)\right)$.

\section{Simultaneous diagonalization \\ in the tunnel limit}
\label{ap:sim_diag}
 %The matrices $\mathbb{C}$ and $\mathbb{J}$ in Eq. \ref{eq:matricesCJ} are positive definite, allowing a transformation fulfilling $C^T\mathbb{C}C=\mathbbm{1}$ and $C^T\mathbb{J}C=D_J$, where $D_J$ is the diagonal matrix with the eigenvalues of the generalized diagonalization $\mathbb{J} v =  \lambda\mathbb{C}v$ \cite{Eremenko2019}. Explicitly, as $\mathbb{C}$ is positive definite, it can be written as

 The matrix $\mathbb{C}$ in Eq. \ref{eq:matricesCJ} is positive definite, hence it can be written \cite{Eremenko2019}
 \begin{equation}
     \mathbb{C} = (v_C) D_C (v_C)^T = \left[(v_C) \sqrt{D_C}\right]\left[(v_C) \sqrt{D_C}\right]^T := R^T R, \nonumber
 \end{equation}

\noindent
where $v_C$ is the matrix made of the orthonormal vectors that diagonalize $\mathbb{C}$ into $D_C$. Note that if we define $C_N{=}R^{-1}U$, with any orthogonal $U^T{=}U^{-1}$, we have $C_N^T\mathbb{C}C_N=\mathbbm{1}$. Now, consider that there is $C_\phi$ such that $C_\phi^T\mathbb{J}C_\phi=D_J$ with diagonal $D_J$, and that $C_\phi^T=C_N^{-1}$ for reasons that will be clear afterwards -- it is allowed because in a Hamiltonian formulation the conjugate variables are independent \cite{Han1999}. Then the previous equation for $D_J$ can be written as the diagonalization $U^T(R\mathbb{J}R^T)U=D_J \leftrightarrow R\mathbb{J}R^TU = U D_J$, showing that $U = (u_1, u_2, ...)$ is made of the eigenvectors from that diagonalization, which is possible because the matrix product between parenthesis is symmetric. Multiplying by $R^{-1}$ it is equivalent to 
\begin{equation}
 \mathbb{J}R^TU = R^{-1} U D_J = R^{-1} R^{-1T} V D_J = \mathbb{C} V D_J, \nonumber  
\end{equation}

\noindent
where $V:=R^TU$ shows the structure of a generalized diagonalization into the columns of $V=(v_1, v_2, ...)$, with eigenvalues $\lambda$ satisfying $\textrm{det}(\mathbb{J}-\mathbb{C}\lambda)=0$. Note that $v_n^T \mathbb{C}^{-1} v_m = v_n^T R^{-1} R^{-1T} v_m = u_n^T u_m = \delta_{nm}$.
 
Then we can write $H^{\textrm{eff}}_{\textrm{harm}} {=} \vec{N}'^T \vec{N}' {+} \vec{\phi}'^T D_J \vec{\phi}'$, with $\vec{N}' {=} C_N^{-1}\vec{N}$ and $\vec{\phi}' {=} C_\phi^{-1}\vec{\phi}$. Each transformed pair $(\hat{N}'_\nu,\hat{\phi}'_\nu)$ is uncoupled from the others, and we have $\comm{\hat{\phi}'_\mu}{\hat{\phi}'_\nu} {=} 0$, $\comm{\hat{N}'_\mu}{\hat{N}'_\nu} {=} 0$ and $\comm{\hat{\phi}'_\mu}{\hat{N}'_\nu} {=} i\sum_\alpha (C^{-1}_\phi)_{\mu\alpha} (C^{-1}_N)_{\nu\alpha} {=} i(C^{-1}_\phi C^{-1T}_N)_{\mu\nu} {=} i\delta_{\mu\nu}$, where the choice in the previous paragraph allows to conserve the commuting properties in the transformed operators.

Now, defining $a_\nu^{(\dagger)} = c_{1\nu}^{(*)}\hat{N}'_\nu + c_{2\nu}^{(*)}\hat{\phi}'_\nu$, imposing the commutation relations $\comm{a_\mu}{a_\nu^\dagger} = \delta_{\mu\nu}$ constrains $c_{1\nu}c_{2\nu} {=} {-}1/2$. Then, $H^{\textrm{eff}}_{\textrm{harm}}$ can be made diagonal with the choice $c_{1,2\nu} {=} (D_J)_\nu^{\pm1/4}i^{-1/2\pm1/2}/\sqrt{2}$. 
%Imposing the last commutator to be equal to $1$, we see that this will be in general possible if $\mathbb{C}$ is diagonal. Then consider this is the case, with $E_{C_3}=0$ (the general case can be solved expanding the number of modes: situation with 4 terminals where $E_{J_{\nu4}}=0$ and $E_{C_4}=0$ (diagonal $\mathbb{C}$), allows to solve it - I think). Then choosing $c_{1,2\,\nu}$'s so that the Hamiltonian is diagonal.  $H^{eff}_{harm} = \sum_{\nu} 2\sqrt{E_{C_\nu}^2 D_{J_\nu}} (a^\dagger_\nu a_\nu + \frac{1}{2})$ with $D_{J_{1,2}} = (a\mp b)/(2E_{C_1}E_{C_2})$ with $a = E_{C_1}(E_{J_{12}}+E_{J_{23}}) + E_{C_2}(E_{J_{12}}+E_{J_{13}})$ and $b = \sqrt{a^2 - 4E_{C_1}E_{C_2}(E_{J_{13}}E_{J_{23}} + E_{J_{12}}(E_{J_{13}}+E_{J_{23}})) }$
Finally, $H^{\textrm{eff}}_{\textrm{harm}} = \sum_{\nu} 2\sqrt{D_{J_\nu}}(a^\dagger_\nu a_\nu + 1/2)$, where $(D_J)_\nu=\left(b\pm\sqrt{b^2-4c}\right)/4$, with
%\begin{align}
%    &b {=} E_{C_1}(E_{J_{12}}+E_{J_{13}}) + 
%    E_{C_2}(E_{J_{12}}+E_{J_{23}}) + E_{C_3}(E_{J_{13}}+E_{J_{23}}) \nonumber \\
%    &c {=} (E_{C_1}E_{C_2}{+}E_{C_2}E_{C_3}{+}E_{C_3}E_{C_1}) (E_{J_{12}}E_{J_{23}}{+}E_{J_{23}}E_{J_{13}}{+}E_{J_{13}}E_{J_{12}}) \nonumber
%\end{align} 
\begin{align}
    &b = \sum_{\alpha=1}^3 E_{C_\alpha}\left(E_{J_{\alpha,\alpha-1}}+E_{J_{\alpha,\alpha+1}}\right) \nonumber \\
    &c = \left( \sum_{\alpha=1}^{3} E_{C_\alpha}E_{C_{\alpha+1}}\right) \left( \sum_{\alpha=1}^{3} E_{J_{\alpha-1,\alpha}}E_{J_{\alpha,\alpha+1}}\right), \nonumber
\end{align}

\noindent
where the indexes values $0$ and $4$ refer to the indexes $3$ and $1$, respectively, and $E_{J_{\mu\nu}} {=} E_{J_{\nu\mu}}$. Some particular cases are $2D_{J_\nu} {=} \{ 0,\, (E_{C_1}{+}E_{C_2})E_{J_{12}}\}$ for $E_{J_{13}}{=}0{=}E_{J_{23}}$, and $2D_{J_\nu}{=}\{ E_{C_1}E_{J_{13}},\, E_{C_2}E_{J_{23}}\}$ for $E_{J_{12}}{=}0{=}E_{C_3}$.\\

\section{Two level island limit}
\label{ap:island2levels}
In the limit $E_{C_3} \gg E_{J_{13}}, E_{J_{23}}$, we can describe island $3$ with its two lowest energy levels corresponding to certain $N_3$ and $N_3+1$, with an energy difference of $2E_{C_3}(n_{g3}-1/2)$. In this case it is useful to describe the other two terminals with $\hat{N}_{12} = \hat{N}_1 - \hat{N}_2$ (note that a third variable is not necessary - the charge conservation imposes the value of the remaining $\hat{N}_1+\hat{N}_2$). The tunneling processes between $1$ and $2$ create an imbalance of two pairs in $N_{12}$ ($-E_{J_{12}}e^{-i2\hat{\phi}_{12}}+h.c.$), while the tunneling from $1$ to the island and from the island to $3$ creates an imbalance of one pair ($-E_{J_{13}}e^{-i\hat{\phi}_{12}} \ket{N_3+1}\bra{N_3} + h.c.$ and $-E_{J_{23}}e^{-i\hat{\phi}_{12}} \ket{N_3}\bra{N_3+1} + h.c.$, respectively). Thus, in the basis $\{ \ket{N_3},\, \ket{N_3+1}\}$, 
\begin{multline}
    H^{\textrm{eff}}_{\textrm{2lvs}} = H_C -E_{J_{12}} \cos{ 2\hat{\phi}_{12}} \, + \\ 
    \begin{pmatrix}
        -E_{C_3}(n_{g_3}-1/2) & -E_{J_{13}}e^{i\hat{\phi}_{12}} - E_{J_{23}}e^{-i\hat{\phi}_{12}} \\
        -E_{J_{13}}e^{-i\hat{\phi}_{12}} - E_{J_{23}}e^{i\hat{\phi}_{12}} & E_{C_3}(n_{g_3}-1/2)
    \end{pmatrix}, \nonumber
\end{multline}

\noindent
where the correspondence with the two-terminal junction with a central level is given by $\xi=E_{C_3}(n_{g_3}{-}1/2)$ and $\Gamma_{1,2} = -E_{J_{13,23}}$ (at $E_{J_{12}}{=}0$). From a different perspective, in the case of $E_{J_{12}} \gg E_{C_{12}}$, this tunneling term can be diagonalized with the charging term into an oscillator coupled to the two-level structure. This is similar to what occurs when coupling a circuit with a reference transmon \cite{Bargerbos2022}.

\bibliography{bib/main.bib}
\end{document}